\documentclass{ws-revol6-5x9-75}
\usepackage{graphicx}

\newcommand{\kket}[1]{\ensuremath{|#1\rangle}}
\newcommand{\bbra}[1]{\ensuremath{\langle #1|}}
\newcommand{\avg}[1]{\ensuremath{\langle#1\rangle}}

\def\draftnote{Copyright\ \copyright\ 2002 by {\bf World Scientific}, reprinted with permission.}

\begin{document}

\title{THE FINITE-TEMPERATURE PATH INTEGRAL MONTE CARLO METHOD AND ITS 
APPLICATION TO SUPERFLUID HELIUM CLUSTERS}

\author{P. Huang$^{\dag}$, Y. Kwon$^{\ddag}$, and K. B. Whaley$^{\dag}$}

\address{$^{\dag}$Department of Chemistry and Kenneth S. Pitzer Center
for Theoretical Chemistry, University of California, Berkeley, CA
94720-1460, USA\vspace{2pt}\\$^{\ddag}$Department of Physics, Konkuk
University, Seoul 143-701, Korea\\ E-mail:
whaley@socrates.berkeley.edu}

\begin{abstract}
We review the use of the path integral Monte Carlo (PIMC) methodology
to the study of finite-size quantum clusters, with particular emphasis
on recent applications to pure and impurity-doped $^4$He clusters.  We
describe the principles of PIMC, the use of the multilevel Metropolis
method for sampling particle permutations, and the methods used to
accurately incorporate anisotropic molecule-helium interactions into
the path integral scheme.  Applications to spectroscopic studies of
embedded atoms and molecules are summarized, with discussion of the
new concepts of local and nanoscale superfluidity that have been
generated by recent PIMC studies of the impurity-doped $^4$He
clusters.\index{Monte Carlo!path integral}
\end{abstract}

\section{Introduction}
\label{sec:intro}
\setcounter{equation}{0}

Over\footnotetext{Preprint from P. Huang, Y. Kwon, and K.~B. Whaley,
in {\em Quantum Fluids in Confinement}, Vol.~4 of {\em Advances in
Quantum Many-Body Theories}, edited by E. Krotscheck and J. Navarro
(World Scientific, Singapore, 2002), in press.} the past 15 years, the
path integral Monte Carlo (PIMC) method has evolved into a uniquely
powerful computational tool for the study of bulk and finite quantum
systems.  In PIMC, one is interested in computing the thermal average
of a quantum observable $\hat{O}$ at a given temperature $T$, which
can be expressed with respect to the thermal density matrix
$\rho(R,R';\beta)=\bbra{R'}e^{-\beta\hat{H}}\kket{R}$:
\begin{equation}
\avg{\hat{O}} = Z^{-1} \int dRdR'\, \rho(R,R';\beta) \bbra{R}\hat{O}\kket{R'},
\label{eq:pathint}
\end{equation}
where $R\equiv ({\bf r}_1,{\bf r}_2,\ldots,{\bf r}_N)$ is a point in
the $3N-$dimensional configuration space of an $N$-particle system,
$\hat{H}$ is the Hamiltonian, and $\beta=1/k_BT$.  Here $Z=\int dR\,
\rho(R,R;\beta)$ is the partition function.  The multidimensional
integral of Eq.~(\ref{eq:pathint}) can in principle be evaluated by
standard Monte Carlo integration schemes, {\em i.e.}\ by taking an
average of $\bbra{R}\hat{O}\kket{R'}$ over the configurations $\{R,
R'\}$ sampled from the probability distribution $Z^{-1}
\rho(R,R';\beta)$.  However, the full density matrix of an interacting
$N$-particle quantum system is generally not known at low
temperatures.  Therefore one needs to resort to the discrete
representation of the Feynman path integral formula for a
low-temperature density matrix, which will be discussed in detail in
Section~\ref{sec:theory}.

The finite-temperature nature of PIMC makes this a complementary
approach to zero-temperature Monte Carlo methods, such as variational
Monte Carlo, or Green's function-based methods such as diffusion Monte
Carlo,\index{Monte Carlo!diffusion} which are reviewed in Chapter~1 of
this volume.~\cite{chapter1} PIMC is currently the only numerical
method capable of directly addressing finite-temperature superfluidity
and the superfluid transition in helium.  In addition, unlike the
zero-temperature methods, PIMC does not require the use of a trial
function, requiring as input only the particle masses, numbers, and
interaction potentials, in addition to the temperature and volume.
Consequently, it is a numerically exact technique, and is independent
of the trial function bias problems that zero-temperature methods may
suffer from.  However, because PIMC provides thermodynamic averages,
state-specific information is generally not available.  Although in
principle the density matrix contains information on the full
eigenspectrum of the Hamiltonian $\hat{H}$, extracting this requires
the numerical inversion of a Laplace
transform.~\cite{pollock84}$^-$\nocite{gallicchio94}\cite{boninsegni96}
Such inversions are known to be notoriously difficult in the presence
of Monte Carlo noise.  Thus, to date, path integral Monte Carlo has
provided only very limited dynamical information of a direct nature.
Nevertheless, it has provided critical microscopic input into
dynamical models for physical systems in helium droplets, and in
conjunction with zero-temperature, state-specific calculations for
these finite helium systems, PIMC has proven to be a powerful means of
investigating the dynamic consequences of atomic scale structure of a
superfluid.~\cite{kwon00}

Feynman first applied the path integral approach to liquid helium in
1953, and provided a consistent clarification of the role of Bose
permutation symmetry in the lambda transition of liquid
helium.~\cite{feynman53} In Feynman's original treatment, the
multidimensional integrals of Eq.~(\ref{eq:pathint}) were approximated
analytically.  In the late 1980's, Ceperley and Pollock subsequently
devised a Monte Carlo scheme for the exact numerical evaluation of
these multidimensional integrals, in which the combined configuration
and permutation spaces were efficiently sampled using a multilevel
Metropolis method.~\cite{pollock87} This allowed direct quantitative
application of Feynman's path integral approach to the superfluid
state of helium for the first time.

Since then, the PIMC method has been applied to provide a quantitative
description of numerous bulk and finite bosonic systems.  In addition
to extensive studies of bulk helium, PIMC has now been employed in the
study of $^4$He/$^3$He mixtures,~\cite{boninsegni95} of helium and
molecular hydrogen droplets,~\cite{kwon00,scharf93} and of helium and
hydrogen films on various surfaces.~\cite{pierce99,gordillo97} In this
work, we focus on the application of PIMC to quantum simulations of
finite helium clusters, $^4$He$_N$.  Since we are primarily concerned
with the bosonic isotope of helium, for the remainder of the chapter,
we will denote $^4$He as simply He, unless explicitly stated.  The
field of helium cluster research has grown very rapidly over recent
years due to new possibilities of inserting molecular probes and
studying their properties.~\cite{toennies01} An overview of the
experimental work in this area is provided in Chapter 9 of this
volume.  Accompanying this rise in experimental studies, there has
been a correspondingly increased demand for complementary theoretical
study of these finite quantum systems.

The earliest PIMC simulation of pure He$_N$ droplets, made in 1989,
demonstrated superfluid behavior for sizes $N$ as small as
64.~\cite{sindzingre89} This result, together with zero-temperature
calculations of the size scaling for pure cluster excitation spectra
made at that time,~\cite{ramakrishna90} was taken to be strong
theoretical evidence that these finite-sized clusters were indeed
superfluid.  This was later confirmed experimentally, through a series
of elegant experiments with impurity-doped helium
clusters.~\cite{hartmann96,grebenev98} Doped clusters present
additional technical challenges for PIMC beyond the requirements posed
by a finite cluster of pure He$_N$.  Early PIMC work with doped
clusters addressed the widely studied He$_N$SF$_6$
system.~\cite{kwon96,kwon99} These studies showed that the global
superfluid fraction appeared not to be significantly modified by
introduction of an impurity.  However, it soon became apparent that
interesting new local features due to Bose exchange symmetry were
present in the immediate vicinity of an impurity.  This led to the
recognition that a local non-superfluid density could be induced by
the molecular interaction with helium.~\cite{kwon00} PIMC simulations
of doped clusters have now been made with a variety of impurities,
including OCS, HCN, benzene, H$_2$, neutral and ionic alkali atoms,
and some complexes of these molecules.  As will be outlined here,
these studies have revealed a broad range of properties of the dopant
as well as insight into the dopant influence on the superfluid
properties of the droplets.  Key features that have emerged from these
studies of doped droplets are the ability to analyze superfluid
behavior in nanoscale dimensions, to characterize quantum solvation in
a superfluid, and to probe the atomic-scale behavior of a superfluid
near a molecular interface.

In this review, we first provide an overview of the Feynman path
integral formalism for quantum statistical thermodynamics in
Section~\ref{sec:theory}.  Following this introduction, we discuss the
PIMC implementation of the general Feynman theory, focusing in
particular on the multilevel Metropolis sampling method of Ceperley
and Pollock.  We then review applications of the PIMC method to the
study of pure helium droplets, and of doped helium clusters containing
atomic or molecular impurities in Section~\ref{sec:app}.  The analysis
of superfluidity in finite droplets, and concepts of nanoscale
superfluids and local superfluidity are described in
Section~\ref{subsec:nsfluid}. PIMC applications to spectroscopic
studies of doped helium clusters are summarized in
Section~\ref{sec:spectra}.  We conclude in Section~\ref{sec:conclude}
with a summary of open questions.

\section{Theory}
\label{sec:theory}
\setcounter{equation}{0}

\subsection{General formulation}

Here we deal with a cluster of $N$ He atoms doped with a single
impurity. In many of the studies made to date, the impurity is assumed
to be fixed at the origin without either translational or rotational
motion.  Neglect of the impurity translational degrees of freedom is
not essential, but for impurity particles which are heavy relative to
a helium atom, it is reasonable and often convenient to ignore the
translational motion of the impurity.  However, the neglect of the
impurity rotational degrees of freedom should be treated with caution,
especially for impurities with small principal moments of inertia.
Incorporation of the rotational motion of the impurity is an area of
current work.  Hence for the present discussion we consider the
following system Hamiltonian $\hat{H}$:
\begin{equation}
\hat{H} = \hat{K} + \hat{V} = - \lambda\sum_{i=1}^{N} \nabla^2_i + \sum_{i<j} 
V_{\mathrm{He-He}}(r_{ij}) + \sum_{i=1}^{N} V_{\mathrm{He-imp}} ({\bf r}_i), 
\label{eq:hamiltonian}
\end{equation}
where $\lambda=\hbar^2/2m_4$, with $m_4$ being the helium mass.  The
potential energy $\hat{V}$ includes a sum of He-He pair potentials,
$V_{\mathrm{He-He}}$, and He-impurity interactions
$V_{\mathrm{He-imp}}$, where the latter are most readily given in the
molecular frame.  If necessary, {\em e.g.} for light molecules, the
impurity translational degrees of freedom can be incorporated by
adding an additional term $-\lambda_I\nabla_I^2$, corresponding to the
impurity center-of-mass kinetic energy.

In the path integral approach, one uses the identity
$e^{-(\beta_1+\beta_2)\hat{H}} = e^{-\beta_1\hat{H}}
e^{-\beta_2\hat{H}}$ to express the low-temperature density matrix by
an integral over all possible {\em paths},
$\{R,R_1,R_2,\ldots,R_{M-1},R'\}$, with the weight for each path given
by the product of density matrices at a higher temperature $T'=MT$:
\begin{equation}
\rho(R,R';\beta) = \int dR_1dR_2 \ldots dR_{M-1}\, \rho(R,R_1;\tau) \rho(R_1,R_2;\tau) \ldots \rho(R_{M-1},R';\tau). \label{eq:convol}
\end{equation}
Here $\tau\equiv\beta/M=(k_BT')^{-1}$ constitutes the imaginary {\em
time step} defining the discrete representation of the path integral.
For a sufficiently high temperature $T'$ or, equivalently, for a small
enough time step $\tau$, there exist several approximations to the
density matrix that are sufficiently accurate for this factorization
to be used in numerical work.~\cite{ceperley95} The simplest of these
high-temperature approximations is the {\em primitive} approximation,
which is based upon the Trotter formula:~\cite{trotter59}
\begin{equation}
\rho(R_k,R_{k+1};\tau) \approx \int dR'\, \bbra{R_k}e^{-\tau\hat{K}}\kket{R'} 
\bbra{R'}e^{-\tau\hat{V}}\kket{R_{k+1}}. \label{eq:prima}
\end{equation}
The potential energy operator $\hat{V}$ in Eq.~(\ref{eq:hamiltonian})
is diagonal in the position representation,
\begin{equation}
\bbra{R'}e^{-\tau\hat{V}}\kket{R_{k+1}} = e^{-\tau V(R_{k+1})} \delta(R'-R_{k+1}), \label{eq:potact}
\end{equation}
while the kinetic term corresponds to the free particle density
matrix:
\begin{equation}
\bbra{R_k}e^{-\tau\hat{K}}\kket{R'} = \rho_0(R_k,R';\tau) = (4\pi\lambda\tau)^{-3N/2} e^{-(R_k-R')^2/4\lambda\tau}. \label{eq:kinact}
\end{equation}
From Eqs.~(\ref{eq:convol})--(\ref{eq:kinact}), the path integral
representation for the density matrix in the primitive approximation
may be expressed as:\index{density matrix!path integral
representation}
\begin{eqnarray}
%\lefteqn{\rho(R_0,R_M;\beta) =} \\
% & & (4\pi\lambda \tau)^{-3NM/2} \int dR_1dR_2 \cdots dR_{M-1}\, 
%\exp\left[-\sum_{k=1}^M \sum_{i=1}^N 
%\frac{({\bf r}_{i,k}-{\bf r}_{i,k-1})^2}{4\lambda\tau} \right. \nonumber \\
%& & \left. - \tau\sum_{k=1}^M V(R_k)\right], \nonumber \label{eq:primitive}
&& \rho(R_0,R_M;\beta) = 
\label{eq:primitive} \\
&& (4\pi\lambda \tau)^{-3NM/2} \int dR_1 \cdots dR_{M-1}\, 
\exp\left[-\sum_{k=1}^M \sum_{i=1}^N 
\frac{({\bf r}_{i,k}-{\bf r}_{i,k-1})^2}{4\lambda\tau}  
 - \tau\sum_{k=1}^M V(R_k)\right], \nonumber 
\end{eqnarray}
with
\begin{equation}
V(R_k) = \sum_{i<j}^N V_{\mathrm{He-He}}(r_{ij,k}) + 
\sum_{i=1}^N V_{\mathrm{He-imp}}({\bf r}_{i,k}). \label{eq:V} 
\end{equation}
A single $R_k$ is referred to as a {\em time slice}, and ${\bf
r}_{i,k}$, the position of the $i^{\mathit th}$ particle at the
$k^{\mathit th}$ time slice, as a {\em bead}.  From here on, $i$ will
denote particle index and $k$ will denote time index.
Eq.~(\ref{eq:primitive}) may be viewed as a classical configuration
integral, with the exponent of its integrand corresponding to an
energy function.  The first term in the exponent, derived from the
kinetic energy, corresponds to a spring potential that connects beads
representing the same atom at successive imaginary times, with
coupling constant $(2\lambda\tau)^{-1}$. This chain of beads connected
by springs is often referred to as a {\em
polymer}.~\cite{feynman53,chandler81} The helium-helium interaction is
represented by an inter-polymer potential that has non-zero
interactions only between beads located on different polymers and
indexed by the same imaginary time value.  This corresponds to
Feynman's original idea of mapping path integrals of a quantum system
onto interacting classical polymers, with a special form of polymer
interaction potential.~\cite{feynman53} In the absence of kinetic
contributions from the impurity, the helium-impurity interaction acts
as an additional external field for the helium atoms, and hence for
these polymers.

The thermal density matrix
$\rho(R,R';\beta)=\bbra{R}e^{-\beta\hat{H}}\kket{R'}$ can be expanded
in terms of the eigenvalues $\{E_n\}$ and the corresponding
eigenfunctions $\{\phi_n\}$ of the Hamiltonian $\hat{H}$:
\begin{equation}
\rho(R,R';\beta) = \sum_n \phi_n(R) \phi^*_n(R') e^{-\beta E_n}. \label{eq:dmatrix}
\end{equation}
This is appropriate for a system of distinguishable particles under
Boltzmann statistics.  For a Bose system such as He or para-H$_2$, it
should be symmetrized with respect to particle exchanges.  This can be
done by modifying the sum in Eq.~(\ref{eq:dmatrix}) to a sum over
exchange-symmetrized stationary states $\tilde{\phi}_{\alpha}$ only:
\begin{equation}
\rho_B(R,R';\beta) = \sum_{\alpha} \tilde{\phi}_{\alpha}(R) \tilde{\phi}^*_{\alpha}(R') e^{-\beta E_{\alpha}}. \label{eq:symmsum}
\end{equation}
A symmetrized eigenfunction $\tilde{\phi}_{\alpha}(R)$ can be obtained
by summing a stationary wavefunction of distinguishable particles,
$\phi_n({\cal P}R)$, over all $N$-particle permutations ${\cal
P}$:~\cite{feynman72}
\begin{equation}
\tilde{\phi}_{\alpha}(R) = \frac{1}{N!} \sum_{\cal P} \phi_n ({\cal P}R). \label{eq:symmwf}
\end{equation}
Inserting Eq.~(\ref{eq:symmwf}) into Eq.~(\ref{eq:symmsum}), we obtain
the symmetrized density matrix for a Bose system:\index{density
matrix!Bose symmetry}
\begin{equation}
\rho_B(R,R';\beta) = \frac{1}{N!} \sum_{\cal P} \rho(R,{\cal P}R';\beta). 
\label{eq:bosedm}
\end{equation}

\subsection{Density matrix evaluation}\index{density matrix!evaluation of}

In order to make PIMC calculations more tractable, one wishes to use
the smallest possible number of time slices $M$ for a given
temperature $T$.  This means that it is essential to find accurate
high-temperature density matrices at as {\em small} a value $T'$ as
possible, so that the imaginary time step $\tau$ be kept as {\em
large} as possible.  It has been found that for the helium-helium
interaction, the primitive approximation described in
Eqs.~(\ref{eq:prima})--(\ref{eq:kinact}) is accurate enough at
temperatures higher than $\sim$1000~K.~\cite{ceperley92} This implies
that a PIMC simulation at $T\sim 0.3-0.4$~K, where the spectroscopic
measurements for the impurity-doped helium clusters have been
performed,~\cite{toennies98} would require several thousands of time
slices. This computational expense can be avoided by going beyond the
primitive approximation to a more sophisticated approximation for the
high-temperature density matrices.  Based on the Feynman-Kac
formula,~\cite{ceperley92} the high-temperature density matrix
$\rho(R,R';\tau)$ can be approximated by a product of the free
particle propagator $\rho_0(R,R';\tau)$ of Eq.~(\ref{eq:kinact}) and
an interaction term $e^{-U(R,R';\tau)}$:
\begin{equation}
\rho(R,R';\tau) \approx \rho_0(R,R';\tau)e^{-U(R,R';\tau)}. \label{eq:highT}
\end{equation}
The interaction term $e^{-U(R,R';\tau)}$ is in turn factored into
contributions deriving from the helium-helium and helium-impurity
interactions, $\rho_{\mathrm{He-He}}$ and $\rho_{\mathrm{He-imp}}$,
respectively.  For spherical interactions, one can generate a
pair-product form of the exact two-body density matrices using a
matrix squaring approach discussed in detail in
Ref.~\refcite{ceperley95}.  We use this for the helium-helium
interaction, which is spherically symmetric.  Such helium-helium
density matrices of the pair-product form have been shown to be
accurate for $\tau^{-1}/k_B \ge 40$~K, {\em i.e.}, $T' \ge
40$~K.~\cite{pollock87,ceperley92} This same approach can be used for
the helium-impurity interaction when this is also isotropic.  However,
for molecules the helium-impurity interaction
$V_{\mathrm{He-imp}}({\bf r})$ is in general not isotropic, and may
involve complicated three-dimensional dependencies.  This can be dealt
with in several ways.  One approach is to expand the helium-impurity
interaction in spherical terms and then employ pair-product forms as
above.  We have found it convenient to work within the primitive
approximation for the helium-impurity interaction, which allows
considerable flexibility when changing impurities.  The required time
step for the accurate primitive helium-impurity density matrices
varies, depending on the impurity molecule involved ({\em e.g.},
$\tau^{-1}/k_B \ge 80$~K for He-SF$_6$ and He-OCS, $\tau^{-1}/k_B \ge
160$~K for He-benzene).  This must be recalibrated, {\em e.g.}, by
establishing converged helium densities, for every new molecule that
is studied.  The same re-calibration requirement holds also for
spherical expansions.

\subsection{Multilevel Metropolis algorithm}

For a diagonal operator $\hat{O}$ in the position representation,
$\bbra{R}\hat{O}\kket{R'} = O(R)\delta(R-R')$, we need to consider
only the diagonal density matrices for evaluation of its thermal
average, Eq.~(\ref{eq:pathint}).  For the diagonal density matrix,
both the sum over permutations in Eq.~(\ref{eq:bosedm}) and the
multidimensional integration in Eq.~(\ref{eq:convol}) can be evaluated
by a sampling of discrete paths which end on a permutation of their
starting positions, {\it i.e.}, $s=\{R_0,R_1,R_2,\cdots,R_{M-1},R_M\}$
with $R_M = {\cal P}R_0$.  This gives rise to an isomorphic mapping
onto {\em ring} polymers.  In fact, all physical quantities discussed
in this review can be estimated from a set of stochastically sampled
ring polymers.

In the sampling process, it will be natural to choose the probability
density function as
\begin{equation}
\pi_s = Z^{-1} \prod_{k=0}^{M-1} \rho(R_k,R_{k+1};\tau). \label{eq:pdf}
\end{equation}
The Metropolis algorithm, a widely-used Monte Carlo sampling
technique, provides a route to obtain the converged distribution
$\pi_s$ in the limit of many sampled configurations, as long as the
detailed balance condition is satisfied for transitions between
successive configurations:
\begin{equation}
\pi_s P_{s \rightarrow s'} = \pi_{s'} P_{s' \rightarrow s}. \label{eq:debalance}
\end{equation}
Here $P_{s \rightarrow s'}$ is the transition probability from a
configuration $s$ to $s'$.  This is factorized into an {\em a~priori}
sampling distribution $T_{s \rightarrow s'}$ and an acceptance ratio
$A_{s \rightarrow s'}$:
\begin{equation}
P_{s \rightarrow s'} = T_{s \rightarrow s'} A_{s \rightarrow s'}. 
\label{eq:transprob}
\end{equation}

\begin{figure}
\begin{center}
\scalebox{0.5}{\includegraphics{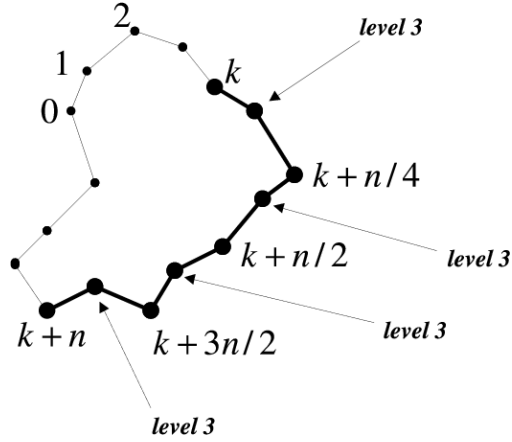}}
\end{center}
\vspace{1pc}
\caption[]{Schematic of multilevel sampling.  The figure shows a
ring polymer of configuration beads for a single particle
corresponding to $M=16$ time slices, to be updated with a three-level
($l=3$) sampling of $2^3 = 8$ time slices simultaneously.  The bold
connections indicate the section of 8 time slices that is to be
updated.}
\label{fig:multilevel}
\end{figure}

In order to speed up convergence times in a path integral simulation,
in particular one involving permutation moves, it is very important to
select an appropriate distribution function $T_{s \rightarrow s'}$ for
a trial move $s'$ from $s$.  The most efficient way of doing this is
the multilevel Metropolis algorithm\index{Monte Carlo!multilevel
Metropolis sampling} developed by Pollock and
Ceperley.~\cite{ceperley95,ceperley92} Here one first chooses end
points of each path by sampling a permutation ${\cal P}$.  Then the
paths are bisected and the configurations at the midpoints
sampled. This process of bisection and midpoint sampling is repeated
multiple times, resulting in a multilevel scheme that samples whole
sections of the paths in a single step.  The acceptance ratio at each
level of this multilevel Markov process is set so that the combined
process of permutation and configuration moves may lead to the
probability density function $\pi_s$ of Eq.~(\ref{eq:pdf}).  Detailed
procedures are summarized as follows:~\cite{ceperley92}
\begin{enumerate}

\item Initialize a configuration $s$.  Typically one starts from a
{\em classical} configuration, in which all beads representing each
atom are located at the same site. So each polymer corresponds
initially to a single point.

\item Choose a time slice $k$ randomly between $0$ and $M-1$ and
construct a table for trial permutation transitions between time
slices $k$ and $k+n$, where $n=2^l$ and $l$ is the {\em level} of this
path updating process.  For the simulation of a He system, $l=3$ turns
out to be a good choice for the permutation moves.  Trial permutations
may be restricted to cyclic permutations among 2, 3, or 4 particles.
The probability for permutation transitions is proportional to
\begin{eqnarray}
T_{\cal P} & = & \exp\left[\frac{-(R_k-{\cal P}R_{k+n})^2}{4\lambda n\tau}\right], \\
C_I & = & \sum_{\cal P} T_{\cal P}. \label{eq:ptable}
\end{eqnarray}
Thus the transition probability for permutational moves does not
depend on the potential energy.  Note that one can explore the entire
$N$-particle permutation space by repeatedly sampling cyclic
permutations among a small number of particles.

\item Select a trial permutation ${\cal P}$ involving $p$ atoms 
such that
\begin{equation}
\sum_{{\cal P'} < {\cal P}} T_{\cal P'} < \chi < \sum_{{\cal P'} \le {\cal P}} T_{\cal P'}, \label{eq:pmove}
\end{equation}
where $\chi$ is a random number on $(0,C_I)$.  This selects the
permutation with probability $T_{\cal P}/C_I$.  Then compute $\Delta_0
= T_{\cal P} / T_I$.  After this, we will sample the intermediate path
coordinates connecting $R_k$ with ${\cal P}R_{k+n}$.  The coordinates
of the $(N-p)$ atoms not on the cycle represented by ${\cal P}$ will
not change from their old positions.  This is level 0 sampling.

\item Start a bisection algorithm by sampling a new midpoint
$R'_{k+\frac{n}{2}}$.  For the sampling distribution function
$T(R'_{k+\frac{n}{2}}|R_k,{\cal P}R_{k+n};\frac{n}{2}\tau)$, we use a
multivariate Gaussian form centered at the mean position $\bar{R}=(R_k
+ R_{k+n})/2$ (see Eq.~(5.16) of Ref.~\refcite{ceperley95}).  Then
compute
\begin{equation}
\Delta_1 = \frac{\rho(R_k,R'_{k+\frac{n}{2}};\frac{n}{2}\tau) \rho(R'_{k+\frac{n}{2}},{\cal P}R_{k+n};\frac{n}{2}\tau)}{\rho(R_k,R_{k+\frac{n}{2}};\frac{n}{2}\tau) \rho(R_{k+\frac{n}{2}},R_{k+n};\frac{n}{2}\tau)}. \label{eq:del1}
\end{equation}
Proceed to the next step with probability
\begin{equation}
\frac{\Delta_1 T(R_{k+\frac{n}{2}}|R_k,R_{k+n};\frac{n}{2}\tau)}{\Delta_0 T(R'_{k+\frac{n}{2}}|R_k,{\cal P}R_{k+n};\frac{n}{2}\tau)}.
\end{equation}
If rejected, go back to step 3 and sample a new trial permutation.
This is level 1 sampling.

\item At the second level, sample $R'_{k+\frac{n}{4}}$ and
$R'_{k+\frac{3n}{4}}$ by bisecting the two intervals and continue to
the next level with the same procedures as used in step 4. This
bisection process is repeated until we get to the final $l$-th level.
At the $l$-th level, sample $R'_{k+1}$, $R'_{k+3}$, $\ldots$, and
$R'_{k+n-1}$ with the probability distribution function $T(R'_j
|R'_{j-1},R'_{j+1};\tau)$.  Proceed to the next step with probability
\begin{equation}
\frac{\Delta_l}{\Delta_{l-1}} \prod_{j=k+1}^{k+n-1} \frac{T(R_j|R_{j-1},R_{j+1};\tau)}{T(R'_j |R'_{j-1},R'_{j+1};\tau)},
\end{equation} 
where
\begin{equation}
\Delta_l = \prod_{j=k+1}^{k+n} \frac{\rho(R'_{j-1},R'_j;\tau)}{\rho(R_{j-1},R_j;\tau)}.
\end{equation}
If rejected, go back to step 3.  Fig.~\ref{fig:multilevel} depicts the
structure of a multi-level sampling with $l=3$, in a path integral
containing $M=16$ time slices.

\item Construct a new permutation table for all 2, 3, or 4 particle
exchanges ${\cal P'}$ acting on ${\cal P}R_{k+n}$.  Accept a new path
$R_k,R'_{k+1},\ldots,R'_{k+n-1},{\cal P}R_{k+n}$ with probability $C_I
/ C_{\cal P}$, where $C_{\cal P} = \sum_{\cal P'} T_{{\cal P'}{\cal
P}}$.  If rejected, go again back to step 3.
      
\item After replacing old coordinates and permutation table with new
ones, repeat steps 3 to 6.
  
\item After attempting several hundreds or thousands of permutation
moves between times slices $k$ and $k+n$, followed by the bisection
procedures to update the midpoints of all $l$ levels, we select a new
time slice $k$ and repeat steps 2 to 6.

\end{enumerate}

One can check that the multilevel bisection algorithm described here
satisfies the detailed balance condition in Eq.~(\ref{eq:debalance}).
Note that $P_{s \rightarrow s'}$ is the total transition probability
to go through all $l$ levels.

\subsection{Estimators for some physical quantities}
\label{subsec:estimators}

With the generalized Metropolis sampling of the permutation
symmetrized density matrix $\rho_B(R,R;\beta)$, the thermal average of
an observable $\hat{O}$ diagonal in the position representation can be
estimated by taking an arithmetic average of
$O(R)=\bbra{R}\hat{O}\kket{R}$ over the paths sampled.  For instance,
the helium density distribution around an impurity molecule can be
estimated by\index{density!path integral estimators}
\begin{equation}
\rho({\bf r}) = \frac{1}{M} \sum_{k=0}^{M-1} \sum_{i=1}^N \avg{\delta({\bf r}-\vec{\bf r}_{i,k})}. \label{eq:density}
\end{equation}
Note that all time slices in ring polymers can be considered as
equivalent.  Unlike importance-sampled diffusion Monte Carlo
methods,\index{Monte Carlo!diffusion} the PIMC calculation of
structural properties such as the density distribution does not
involve any trial function bias.

There are many ways to compute the energy in PIMC, discussed in detail
in Ref.~\refcite{ceperley95}.  Most of the PIMC applications discussed
here employ the {\em direct} estimator obtained by directly applying
the Hamiltonian operator to the density matrix in the position space.
For calculations neglecting the impurity translational and rotational
degrees of freedom, the kinetic energy average is expressed in the
path integral representation as
\begin{eqnarray}
\avg{K} & = & \left\langle \frac{1}{M} \sum_{k=0}^{M-1} \left[\frac{3N}{2\tau} - \frac{(R_k-R_{k+1})^2}{4\lambda\tau^2}  \right.\right. \\
        &   & \left.\left. - \frac{(R_k-R_{k+1})\cdot\nabla_k U^k}{\tau} + \lambda\nabla_k^2 U^k - \lambda(\nabla_k U^k)^2\right] \right\rangle, \label{eq:energy} \nonumber
\end{eqnarray}
where $\nabla_k = \partial / \partial R_k $ is the $3N$-dimensional
gradient operator, and $U^k \equiv U(R_{k-1},R_k;\tau)$ is the
interaction for link $k$, {\em i.e.}, for the spring connecting beads
$k-1$ and $k$ (see Eq.~(\ref{eq:highT})).  Computation of the
potential energy is straightforward since this is diagonal in the
position representation:
\begin{equation}
\avg{V} = \frac{1}{M} \sum_{k=0}^{M-1} \avg{V(R_k)}. \label{eq:penergy}
\end{equation}
As noted earlier, the inter-polymeric potential acts only between
beads defined at the same imaginary time.

One of the most interesting properties of bulk and finite He systems
is their superfluid behavior.  For bulk systems superfluid estimators
are generally derived from linear response theory, {\em i.e.}\ by
considering the helium response to boundary motion.~\cite{baym69}
Pollock and Ceperley showed how to derive momentum density correlation
functions that quantify the superfluid response of bulk systems with
periodic boundary conditions.~\cite{pollock87} Sindzingre {\em et
al.}\ subsequently developed a global linear response
estimator\index{superfluidity!global} for finite helium clusters with
free boundaries.~\cite{sindzingre89} This estimator is based on the
response to a rotation of continuous angular frequency, {\em i.e.} to
a classical rotation such as might be appropriate to a macroscopic
droplet.  Consider the Hamiltonian in a coordinate frame rotating
about an axis with frequency ${\bf\omega}$,
\begin{equation}
\hat{H}_{\mathrm{rot}} = \hat{H}_0 - \hat{\bf L}\cdot{\bf \omega},
\end{equation}
where $\hat{\bf L}$ is the total angular momentum operator.  For a
classical fluid, in the limit of an infinitesimally small rotation the
entire fluid should rotate rigidly with classical moment of inertia
$I_{cl}$.  But in a Bose superfluid, only the normal component
responds to the rotation, resulting in an effective moment of inertia
\begin{equation}
I = \left.\frac{\partial\langle \hat{\bf L}\rangle}{\partial\omega}\right|_{\omega=0}.
\end{equation}
Note that this is to be evaluated in the limit $\omega\rightarrow 0$,
appropriate to rotation of a macroscopic system.  In a homogeneous
system the normal fraction can be defined as
\begin{equation}
\frac{\rho_n}{\rho} = \frac{I}{I_{cl}},
\end{equation}
and the complementary superfluid
fraction\index{superfluidity!fraction} is then given by
\begin{equation}
\frac{\rho_s}{\rho} = 1 - \frac{\rho_n}{\rho} = \frac{I_{cl}-I}{I_{cl}}. 
\label{eq:frac_I}
\end{equation}
For ${\bf\omega}=\omega\hat{z}$, this linear response
estimator\index{superfluidity!path integral estimators} can be
expressed in the path integral representation as~\cite{sindzingre89}
\begin{equation}
\frac{\rho_s}{\rho} = \frac{4m^2\langle A_z^2 \rangle}{\beta\hbar^2 I_{cl}}, 
\label{eq:area_estimator}
\end{equation}
with the vector quantity ${\bf A}$ defined as
\begin{equation}
{\bf A} = \frac{1}{2}\sum_{i,k}({\bf r}_{i,k}\times{\bf r}_{i,k+1}).
\end{equation}
Here the summation runs over particle $i$ and imaginary-time slice
$k$.  The vector quantity ${\bf A}$ is the directed total area of
closed imaginary-time polymers spanned by all $N$ particles, {\it
e.g.}, $A_z$ is the projection of ${\bf A}$ on the
$\hat{z}$-direction.  The average size of a single polymer is given by
its thermal de Broglie wavelength
$\Lambda=(\beta\hbar^2/m)^{1/2}$. This becomes negligible in the
high-temperature limit, and thus the corresponding superfluid
fraction\index{superfluidity!fraction} $\rho_s/\rho$ goes to zero
(although the projected area can remain finite at the microscopic
level).  At low temperatures, when the de Broglie wavelength becomes
comparable to the inter-polymer spacing, particle exchanges cause
polymers to cross-link and form larger ring polymers.  The projected
area ${\bf A}$ increases correspondingly and the helium system attains
an appreciable superfluid fraction.

We discuss the application of this finite-system superfluid estimator
in detail in Sec.~\ref{sec:app}, together with analysis of how this
can be decomposed into local contributions for an inhomogeneous
system.~\cite{kwon00,kwon99,draeger01}

\section{Superfluidity and quantum solvation of atoms and molecules in
bosonic helium clusters}
\label{sec:app}
\setcounter{equation}{0}

Spectroscopic studies of impurity-doped clusters have allowed
experimental investigation of a variety of excitations in helium
clusters.~\cite{toennies01} The relevant temperature range currently
accessible is $T\sim 0.15-0.5$~K.~\cite{toennies98} Thus, the
incorporation of Bose symmetry is essential in simulation of these
systems.  In this section, we focus on the application of
finite-temperature PIMC to bosonic helium clusters.  We begin by
briefly reviewing studies of pure clusters in
Sec.~\ref{wh_subsec:pure}, and then focus on the more recent work for
the clusters doped with various impurities in
Secs.~\ref{subsec:atomic}--\ref{subsec:nsfluid}.
Secs.~\ref{subsec:atomic}--\ref{subsec:mol_impurity} summarize the
structural and energetic aspects, while Sec.~\ref{subsec:nsfluid}
deals with the microscopic analysis of superfluid properties of the
doped clusters.

\subsection{Pure clusters}\index{clusters!pure}
\label{wh_subsec:pure}

Most of the previous theoretical studies involving pure clusters are
based on zero-temperature methods, and have focused on the cluster
elementary excitation spectrum, which qualitatively retains the
phonon-roton features characteristic of bulk He~II\@.  Current work in
this area aims to understand the physical nature of the roton
excitations.~\cite{reatto96} Zero- and finite-temperature calculations
for pure helium clusters have been reviewed
previously,~\cite{whaley94} and so we shall provide only a brief
outline of the finite-temperature results here.  The first studies
were made by Cleveland {\em et al.}, using the path integral molecular
dynamics approach in which exchange is neglected.~\cite{cleveland89}
This allowed structural analysis, which was used to study the changes
in droplet density and diffuseness as a function of size.  Permutation
exchange symmetry was incorporated by Sindzingre {\em et al.}\ in a
study of the temperature and size dependence of the global superfluid
fraction\index{superfluidity!fraction} in finite He$_N$
clusters.~\cite{sindzingre89} These calculations employed the area
estimator discussed in Sec.~\ref{subsec:estimators} and showed that a
broad transition to a predominantly superfluid state occurs at a
temperature depressed from the bulk superfluid transition temperature,
in accordance with expectations from scaling of phase transitions for
finite systems.\index{superfluidity!finite-size} The extent of
depression increased as the cluster size decreased.  For $N=64$, the
onset of the transition occurs just below $T=2$~K, and the transition
appears complete at $T=0.5$~K, with about 90\% or more of the cluster
being superfluid at that temperature.  A qualitative examination of
the relative contribution of long exchange path lengths to the density
revealed that the long exchange path contribution was largest in the
interior of the droplets.

\subsection{Atomic impurities}\index{clusters!doped}\index{impurities!atomic}\index{impurities in He!atomic}
\label{subsec:atomic}

Neutral and ionic atomic impurities constitute the simplest dopants.
For ground electronic states, the helium-impurity interatomic
potential can be calculated with fairly high accuracy using standard
quantum chemistry methods, and the helium-helium interatomic potential
is well-known.  Thus, within the two-body approximation, it is
possible to construct accurate potential energy surfaces for the
ground electronic state.  The interactions of excited electronic
states with helium are, by comparison, less well-characterized and
only a few calculations of electronically excited potential energy
surfaces have been even attempted.  To date, PIMC calculations have
been made for the neutral alkali metal impurities Li, Na,
K,~\cite{nakayama01} and for the ionic impurity
Na$^+$.~\cite{nakayama00}\index{impurities in He!Na, Na$^+$} In
general, the solvation characteristics of each impurity are controlled
by a balance between different energetic
factors.~\cite{ancilotto95,whaley98} These include the helium-impurity
interaction strength, the helium-helium interaction strength, the
impurity kinetic energy (and thus impurity mass), and the free energy
change due to the loss of exchange energy for helium atoms adjacent to
the impurity.

The He-Li, He-Na, and He-K ground state potentials typically have well
depths of $\sim 1-2$~K,~\cite{ancilotto95} smaller than the He-He well
depth of $\sim 11$~K.~\cite{aziz87} By considering these potential
energy factors alone, one would qualitatively expect that the atomic
impurities would reside on the droplet surface in order to minimize
the total energy.  The PIMC studies, made at $T=0.5$~K, indicate that
the neutral alkali impurity species are indeed surface-attached for
cluster sizes of $N\leq 300$.~\cite{nakayama01} In all cases, the
perturbation on the cluster structure due to the presence of the
impurity is weak.  The neutral impurity atom induces small but
distinct modulations in the helium density, starting at the surface
and decaying into the interior of the droplet.  While no
zero-temperature microscopic calculations of these systems have been
made to date, it is expected that this behavior would persist to lower
temperatures and is therefore also applicable to the experimental
studies of these alkali-doped systems that are made at
$T=0.38$~K.~\cite{callegari98}

\begin{figure}
\begin{center}
\scalebox{0.65}{\includegraphics{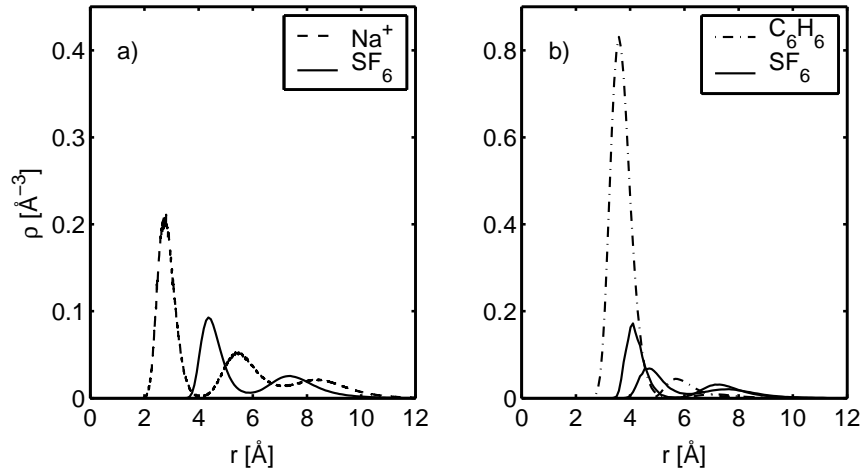}}
\end{center}
\vspace{1pc}
\caption[]{Helium density profiles relative to the impurity center, at
$T=0.625$~K.  The left panel (a) shows the radial density profiles for
He$_{100}$Na$^+$ (dashed lines) and He$_{64}$SF$_6$ (solid lines),
where an isotropic He-SF$_6$ interaction was used.  The right panel
(b) shows anisotropic helium densities for SF$_6$ (solid lines) from a
calculation using an anisotropic He-SF$_6$ interaction,~\cite{kwon99}
viewed along the molecular $C_3$ (higher values) and $C_4$ (lower
values) axes.  This is compared to anisotropic radial density profiles
for He$_{39}$-benzene (dotted-dashed lines),~\cite{kwon01} viewed
along the molecular $C_6$ axis (higher peak) and along the C--H bond
(lower peak).  The He$_{100}$Na$^+$ profile is reproduced from
Ref.~\refcite{nakayama00}.}
\label{fig:denprofile}
\end{figure}

The ionic impurity Na$^+$\index{impurities!charged} interacts much
more strongly with helium and consequently gives a markedly stronger
structural perturbation of the local helium density. The He-Na$^+$
well depth is 407~K, about 40 times larger than the He-He well depth.
Here the finite-temperature PIMC studies indicate that the Na$^+$ ion
resides in the center of the cluster, and the strength and range of
the He-Na$^+$ interaction induces a tightly packed helium ``snowball''
around the ion.~\cite{nakayama00} In Fig.~\ref{fig:denprofile}a, the
radial helium density profile for the He$_{100}$Na$^+$ system at
$T=0.625$~K is shown.  This is compared with the $N=64$ radial density
profile around the molecular impurity SF$_6$, at the same temperature
using an isotropic He-SF$_6$ interaction.  There is a very strongly
modulated layer structure around the Na$^+$ ion, with a high first
coordination shell peak followed by a second peak of lower density.
Similar structural features have been seen in variational shadow
function calculations for Na$^+$ and K$^+$ in bulk
He,~\cite{buzzacchi01,duminuco00} although quantitative differences
exist in comparison with those results.  In the variational
calculations the local angular ordering within the coordination shells
was also examined, leading to more conclusive evidence of solid-like
structure in the first two shells.  These studies indicate that there
definitely exists a more strongly layered shell structure in the
helium density around an impurity ion than around neutral atomic
species, with more solid-like character.  This feature can be further
explored in the imaginary-time path integral representation by
examining the permutation exchanges of helium atoms at specific
locations.  For the He$_{100}$Na$^+$ cluster at $T=0.8$ and $1.25$~K,
the atoms in the first coordination shell rarely participated in
permutations with other particles, and thus are well-localized in the
PIMC sense.  In the second solvation shell, some atoms are involved in
long exchanges at the lower temperature, while in the outer third
shell most atoms are involved in long exchanges.  From this, it was
inferred that the third shell is superfluid, while the second shell
has an intermediate, temperature-dependent
character.~\cite{nakayama00} Such an analysis has been made in more
quantitative detail for the molecular impurities, which we discuss
next.

\subsection{Molecular impurities}\index{cluster!doped}\index{impurities!molecular}
\label{subsec:mol_impurity}

\begin{figure}
\begin{center}
\scalebox{0.5}{\includegraphics{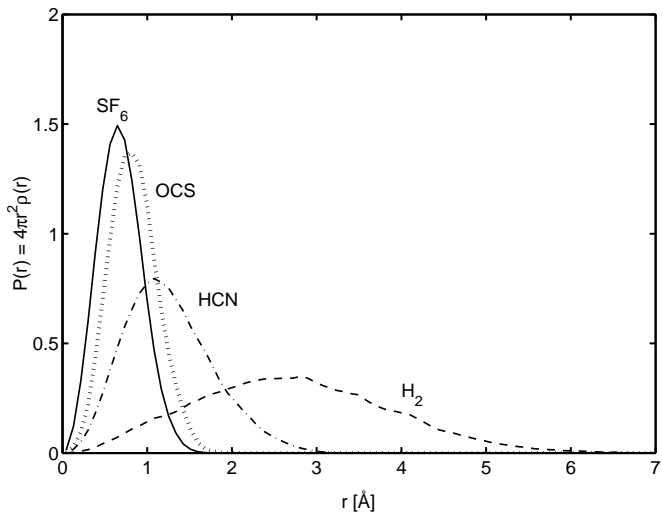}}
\end{center}
\vspace{1pc}
\caption[]{Radial distribution for several impurity
molecules (H$_2$, HCN, SF$_6$, and OCS) relative to the cluster
center-of-mass, shown as $P(r)=4\pi r^2 \rho(r)$ such that $\int
P(r)\, dr = 1$.  All calculations were made from PIMC at $T=0.312$~K,
and include the impurity translational kinetic energy.  Isotropic
interaction potentials were used, and Bose permutation symmetry was
not included.  The H$_2$ and HCN distributions were obtained from a
calculation with $N=128$ He, while the SF$_6$ and OCS distributions
correspond to a cluster of $N=100$ He.  Data courtesy of
D.~T.~Moore.\index{impurities in He!H$_2$}}
\label{fig:displ}
\end{figure}

Molecular impurities introduce an additional level of complexity
because molecules have internal structure and usually possess an
anisotropic interaction with helium.  Especially for the larger
molecules, there is a severe lack of accurate two-body molecule-helium
interaction potentials.  Nevertheless, the study of molecular
impurities in helium clusters is currently of great interest, with an
increasing number of experiments being performed on a variety of
molecules.  Even with simple models for the molecule-helium
interaction, analysis of these experiments in terms of the
perturbation of the helium environment on the molecular internal
degrees of freedom has provided much insight into the quantum fluid
nature of these clusters.  It is important to recall that to date, all
PIMC work involving {\em molecular} impurities in helium have not
explicitly incorporated the impurity rotational kinetic energy.  This
is not an essential restriction, and has been so far made for
convenience rather than for any fundamental limitations.  Since
zero-temperature DMC calculations have recently shown that the helium
densities around small molecules may be sensitive to the rotational
motion of the
molecule,~\cite{kwon00,blume99}$^-$\nocite{viel01}\cite{viel01b} it
would be desirable to incorporate the rotational degrees of freedom in
future path integral studies.  Only for the heaviest rotors can
molecular rotation be justifiably omitted.  Several studies have also
justified neglecting the translational motion of the molecular
impurity, in which case the helium atoms may be regarded as moving in
the external potential field of the molecule in the molecular
body-fixed frame, given by the Hamiltonian of
Eq.~(\ref{eq:hamiltonian}).~\cite{kwon00,kwon01} The validity of this
assumption can be assessed with the comparative study of molecular
delocalization as a function of molecular mass and binding energy
shown in Fig.~\ref{fig:displ}.~\cite{moore01} There, the probability
of finding a molecule at some distance $r$ from the cluster
center-of-mass is shown for a series of molecules.  As expected, the
heaviest dopants such as SF$_6$ and OCS are well-localized near the
center of the cluster, and thus it is a reasonable approximation to
neglect their translational motion.  On the other hand, H$_2$ is much
more delocalized throughout the interior of the cluster, due to its
comparatively smaller mass and weaker helium-impurity interaction, and
therefore requires that its translational motion be properly
incorporated.

\begin{figure}[ht]
\begin{center}
\scalebox{0.8}{\includegraphics{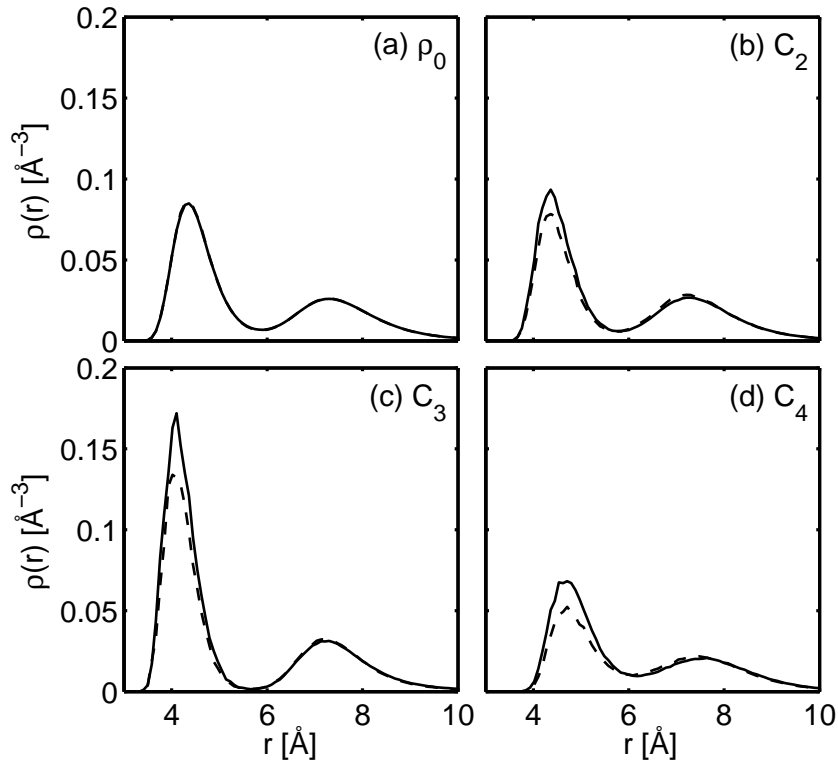}}
\end{center}
\vspace{1pc}
\caption[]{Comparison of the helium density distribution around the
octahedral SF$_6$ molecule in a $N=64$ cluster at $T=0.625$~K and
$T=0.312$~K.  Solid lines show the lower temperature densities and
dashed lines the higher temperature densities.  Panel (a) shows the
angular-averaged density $\rho_0$, for which the profiles at two
different temperatures are identical.  Panels (b), (c), and (d) show
the densities along the three symmetry axes of the molecule $C_2$,
$C_3$, and $C_4$, respectively.  The higher temperature profiles show
consistently smaller peak values in the first solvation shell,
indicating a decrease in the anisotropy of the distribution as
temperature increases.}
\label{fig:tempdep}
\end{figure}

\begin{figure}[ht]
\begin{center}
\scalebox{0.735}{\includegraphics{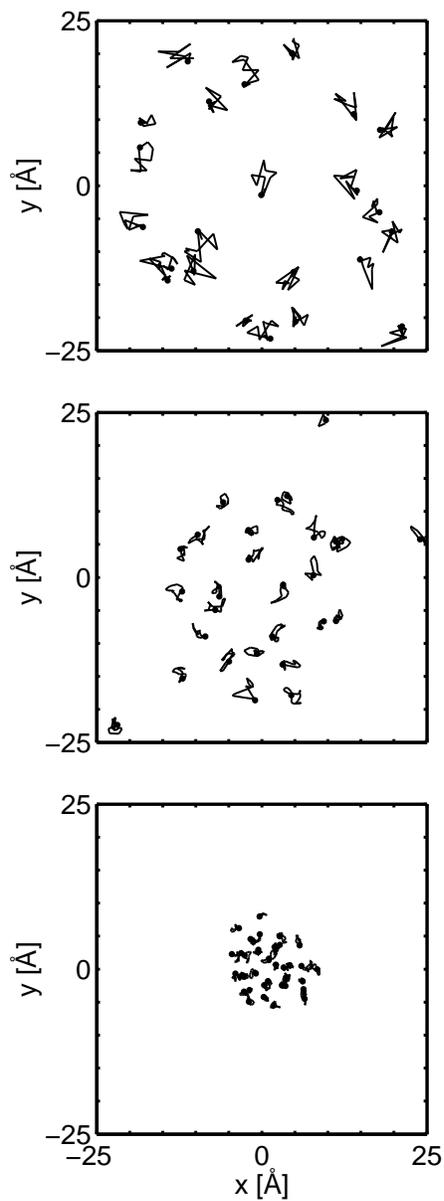}}
\end{center}
\vspace{1pc}
\caption[]{Helium evaporation for the SF$_6$He$_{39}$ cluster.  The
lower panel plots a snapshot of imaginary time paths at $T=1.25$~K. At
this temperature the helium atoms are bound to the cluster.  In the
middle panel, at $T=2.5$~K, the cluster begins to dissociate, loosing
helium atoms.  In the upper panel, at $T=5.0$~K, the cluster has
completely evaporated.}
\label{fig:evaporation}
\end{figure}

To date, the most extensively studied molecular impurity is the
octahedral SF$_6$ molecule.\index{impurities in He!SF$_6$} Early PIMC
work on SF$_6$ in helium clusters employed isotropic molecule-helium
interaction potentials,~\cite{kwon96} and was later extended to
include anisotropic interactions.~\cite{kwon99} The helium structure
around SF$_6$ from an isotropic calculation is shown in
Figs.~\ref{fig:denprofile}a and \ref{fig:sf6iso}. The anisotropic
He-SF$_6$ potential surface has a global minimum of
$-84$~K,~\cite{pack84} considerably deeper than that of the He-He
interaction.  Thus, SF$_6$ is expected to reside at the center of the
cluster.  This has been verified by both zero-temperature
DMC~\cite{mcmahon96} and by finite-temperature PIMC calculations.  For
the He$_{64}$SF$_6$ cluster in the temperature range of
$T=0.3-0.75$~K, there is an anisotropic layering of the helium density
around the SF$_6$.  Integration of the helium density over the first
solvation shell yields about 23 atoms, independent of whether
isotropic or anisotropic interactions are employed.~\cite{kwon00} The
strength and range of the molecule-helium interaction pins the helium
density in the first solvation shell to a total density comparable to
that of the more strongly bound He$_N$-Na$^+$ system.  Detailed
analysis of the helium density distribution around the molecule shows
that while the angular average of the density in the first solvation
shell is independent of temperature below $T=1.25$~K, there is a small
increase in the extent of anisotropy as the temperature is lowered.
This is illustrated in Fig.~\ref{fig:tempdep}, with a comparison of
the densities along different molecular symmetry axes for an $N=64$
cluster at temperatures $T=0.625$~K and $T=0.312$~K.  As the
temperature is increased above $T=1.25$~K, this trend to an
increasingly isotropic distribution is further modified by the onset
of evaporation of helium atoms.  Evaporation begins with atoms in the
second solvation shell, is clearly evident at $T=2.5$~K, and is
essentially complete at $T=5.0$~K (Fig.~\ref{fig:evaporation}).

Since the first experiments for doped clusters that employed SF$_6$ as
a probe species,~\cite{goyal92} a broad array of molecular impurities
have been studied by spectroscopic means.~\cite{toennies98} The
infrared spectral regime has provided a particularly rich field of
study.  Vibrational spectra in the infrared at $T\sim 0.4$ K show
rotational fine structure in $^4$He droplets, but not in $^3$He
droplets, providing evidence that quantum statistics play an important
role in the spectral properties of the dopant.  There is now an
increasing collection of experimental data available for the
rotational dynamics of molecules possessing varying symmetries and a
range of values for the gas phase rotational constant.  To date, PIMC
has been used to make theoretical studies of the linear rotors
OCS~\cite{kwon00,kwon01b} and HCN,~\cite{kwon00} the planar aromatic
molecule benzene (C$_6$H$_6$),~\cite{kwon01} the linear (HCN)$_3$
complex,~\cite{draeger01} and the OCS-(H$_2$)$_M$
complex.~\cite{kwon01b} From these studies the notion of two different
dynamical regimes has emerged, namely that of heavy molecules such as
SF$_6$ that are characterized by gas phase rotational constants $B_0 <
0.5$~cm$^{-1}$, and a complementary regime of lighter molecules
possessing larger gas phase values of $B_0$.~\cite{kwon00} This
division into two dynamical regimes based on rotational constants
emerges from analysis of the helium solvation density and energetics
derived from path integral calculations.

\begin{figure}
\begin{center}
\includegraphics{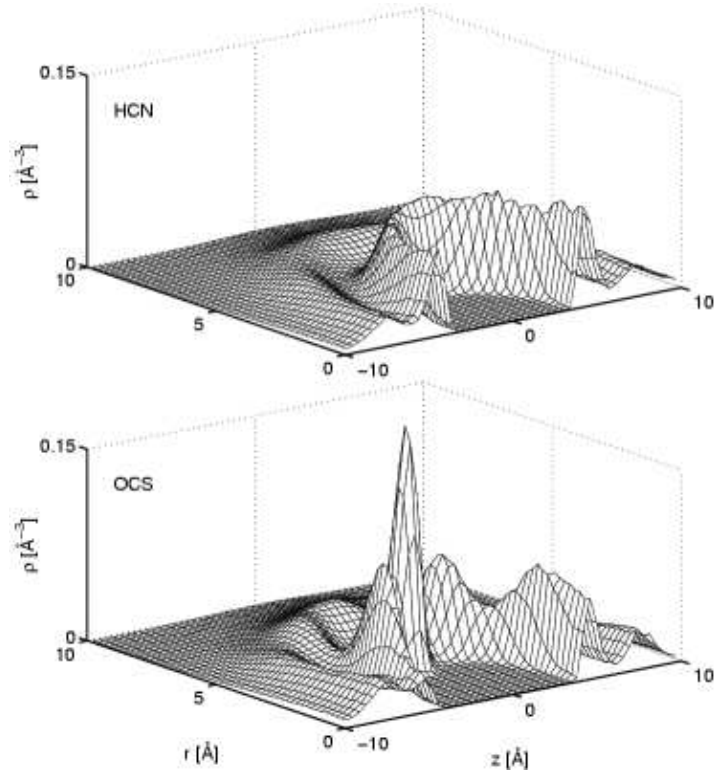}
\end{center}
\vspace{1pc}
\caption[]{Total helium density around HCN (top panel) and OCS (bottom
panel) for a $N=64$ cluster at $T=0.312$~K.~\cite{kwon00} The origin
is set at the impurity center-of-mass.  The OCS is oriented with the
oxygen end directed towards the -$\hat{z}$ direction, and the HCN is
oriented with the nitrogen end directed towards the -$\hat{z}$
direction.}
\label{fig:ocs_hcn_den}
\end{figure}

The OCS impurity\index{impurities in He!OCS} lies in the regime of
relatively heavy molecules, with $B_0=0.20$~cm$^{-1}$.  The He-OCS
potential has a global minimum of $\sim 64$~K,~\cite{higgins99} which
is only about two-thirds that of the He-SF$_6$
potential.~\cite{pack84} It is important to consider the anisotropy of
the intermolecular potential in addition to its strength when
assessing the quantum solvation structure.  In this respect the linear
OCS molecule has lower symmetry than the octahedral SF$_6$.  The
minimum angular barrier for rotation of the OCS about an axis
perpendicular to the molecular axis ({\em i.e.}, the angular adiabatic
barrier for rotation) is 41.9~K.~\cite{higgins99} This barrier is
markedly higher than the corresponding value 20.7~K for SF$_6$, and
consequently gives rise to stronger angular modulations in the
solvating density.~\cite{kwon00} As shown in
Fig.~\ref{fig:ocs_hcn_den}, PIMC calculations for the He$_{64}$OCS
cluster at $T=0.312$~K reveal a strongly structured helium density,
forming approximately elliptical solvation shells around the OCS
impurity.  The first shell integrates to $\sim 17$
atoms.~\cite{kwon00} Because of the axial symmetry of the He-OCS
potential, the density at the global minimum forms a ring around the
OCS molecular axis, consisting of about 6 helium atoms.
 
The benzene molecule (C$_6$H$_6$)\index{impurities in He!benzene} also
lies in the heavy regime.  The benzene $\pi$-electron character leads
to a highly anisotropic interaction with helium, with two deep,
equivalent global potential minimum located on the six-fold axis of
symmetry above and below the plane of the
molecule.~\cite{kwon01,hobza92} A PIMC study of benzene-doped clusters
has shown a highly anisotropic helium structure around the impurity
molecule that reflects this six-fold symmetry.~\cite{kwon01} The
sharpest density peak is located along the C$_6$-axis, at the two
equivalent locations of the global potential minima.  These two global
density maxima are higher than the local density maxima viewed along
the in-plane directions by more than a factor of four, reflecting the
marked anisotropy of the He-benzene interaction potential.  The
extreme density anisotropy is summarized in Fig.~\ref{fig:denprofile}b
where the dotted-dashed lines show density profiles along the
C$_6$-axis and along one of the in-plane directions.  Integration over
any one of the two equivalent global density maxima gives exactly one
helium atom.  We see an interesting effect of near complete
localization of these two helium atoms located at the two global
potential minima on either side of the molecular plane.  As noted in
Ref.~\refcite{kwon01}, this phenomenon can be viewed as a precursor
form of helium adsorption onto a molecular nanosubstrate.  Extending
these studies to larger polyaromatic molecules will allow contact to
be made with PIMC studies of helium adsorption on
graphite.~\cite{pierce99}

In contrast to this highly structured quantum solvation observed
around the heavier molecules such as OCS and benzene, the linear HCN
molecule\index{impurities in He!HCN} falls into the light molecule
regime, with a significantly larger gas phase rotational constant.
For HCN, $B_0=1.48$~cm$^{-1}$.  The He-HCN potential~\cite{atkins96}
is both weaker (its global minimum is $-42$~K) and less anisotropic
than the He-OCS potential.  While there is clearly still an
ellipsoidal layering of the helium density around the HCN, within each
solvation shell there is now a noticeable lack of angular structure,
in contrast to the situation with OCS (Fig.~\ref{fig:ocs_hcn_den}).
For such a light rotor, neglect of the molecular rotational kinetic
energy now becomes a more serious concern.  From DMC studies assessing
the effect of molecular rotation,~\cite{kwon00,viel01b,paesani01} the
expectation here is that the helium density will become more diffuse
when molecular rotation is explicitly incorporated into PIMC.

Self-assembled linear chains of polymeric (HCN)$_M$\index{impurities
in He!(HCN)$_3$} have been detected experimentally in helium
droplets.~\cite{nauta99} The helium structure around such linear
chains has recently been addressed with a study of the properties of
helium droplets with up to $N=500$ atoms that contain (HCN)$_2$ dimers
and (HCN)$_3$ trimers.~\cite{draeger01} Like the monomeric molecules
discussed above, the HCN polymers are found to be located at the
center of the droplet and to induce a layering of the helium density.
Draeger {\em et al.}\ have analyzed the structure of the first
solvation shell around the linear polymer in terms of a
two-dimensional film, estimated the effective confinement potential
for displacement away from the droplet center, and made calculations
for vortex formation in these droplets.~\cite{draeger01} It has been
suggested earlier that the presence of a linear impurity species might
stabilize the formation of a vortex\index{vortices} line in helium
droplets.~\cite{close98} The expectation here is that a vortex line
could be pinned along the molecular axis of a linear molecule such as
HCN, or more likely, along the axis of a linear polymeric chain such
as (HCN)$_M$.  While the physics of vortices constitute an essential
feature of bulk He~II,~\cite{donnelly91} and ways of producing and
detecting vortices during helium droplet formation have been the
subject of much discussion (see Ref.~\refcite{close98} and therein),
no experimental evidence has been found so far for existence of
vortices in finite helium droplets.  Theoretically, vortices have been
shown to be unstable in pure droplets,~\cite{bauer95} and the
situation with regard to doped droplets is still controversial.  The
energy for formation of a vortex, $\Delta E_V$, is defined as
\begin{equation}
\Delta E_V = E_V - E_0,
\end{equation}
where $E_0$ is the ground state cluster energy and $E_V$ is the energy
of the cluster with a vortex line present.  Within the fixed-phase
approximation, the PIMC estimate for this vortex formation energy is
$\sim 30$~K for a He$_{500}$(HCN)$_M$ cluster at $T=0.38$~K, where
$M=0-3$.~\cite{draeger01} In this case, the vortex formation energy is
found not to be significantly affected by the presence of a linear
impurity.  In comparison, density functional calculations made for a
range of impurities and cluster sizes give values of $\Delta E_V$ that
are larger than the fixed-phase PIMC estimates by a factor of 3, and
that are reduced by $\sim 5-10$~K in the presence of an
impurity.~\cite{dalfovo00} An exact estimator for the energy of a
cluster in an angular momentum state $m$ relative to $m=0$ has been
derived using angular momentum projection methods.~\cite{draeger01}
Application of this estimator at $T=2.0$~K indicates that the presence
of an impurity actually results in a slight {\em increase} in the
vortex formation energy.  More work is required in this direction, in
particular the systematic examination of the cluster size and
temperature dependence of $\Delta E_V$ obtained from the angular
momentum projection estimator.

Many other complexes have now been synthesized in helium
droplets.~\cite{toennies01} Indeed, these droplets are proving to be a
remarkably versatile quantum matrix environment for synthesis of
unusual or metastable aggregates.  Of particular interest from a
fundamental point of view are the complexes of OCS with molecular
hydrogen, $H_2$.  Recent spectroscopic measurements on OCS(H$_2$)$_M$
complexes\index{impurities in He!H$_2$} inside He$_N$
clusters\index{clusters!mixed $^3$He/$^4$He} have shown an interesting
feature that has been interpreted as evidence of nanoscale hydrogen
superfluidity.~\cite{grebenev00} Initial PIMC studies of these systems
have been carried out~\cite{kwon01b} using accurate pair potentials of
OCS with He and with H$_2$.~\cite{higgins99,higgins01} Since the
H$_2$-OCS potential surface has a similar angular modulation as that
for He-OCS, but a deeper minimum, the OCS molecule is expected to bind
preferentially to H$_2$ over He.  Calculations for the OCS(H$_2$)$_5$
complex in the He$_{39}$ cluster~\cite{kwon01b} showed that
approximately six helium atoms, which would normally occupy the region
of the global potential minimum in the absence of $H_2$, are
completely displaced by five $H_2$ molecules.  These $H_2$ molecules
form a complete ring encircling the linear OCS molecule at the region
of lowest potential energy.  The helium density is pushed either to
either the secondary peaks in the first shell, or outwards from first
to second shell region.

\subsection{Exchange permutation analysis and impurity-induced 
non-superfluidity}
\label{subsec:nsfluid}

In addition to providing structural and energetic information, PIMC is
currently the only numerical method capable of providing information
on finite-temperature superfluidity in He systems.  At high
temperatures an $N$-body system may be described by Boltzmann
statistics, {\em i.e.}\ in the path integral representation, only the
identity permutation is important.  At low temperatures however,
permutations must be included in the path integral representation for
the thermal density matrix.  In particular, for liquid helium near the
lambda transition, Feynman qualitatively showed that the presence of
long exchange cycles gives rise to the sharp increase in the heat
capacity, but due to the analytical approximations made in his
analysis he was not able to correctly identify the order of the
transition.~\cite{feynman53} Further refinements in this and numerical
PIMC simulations have quantitatively confirmed both the transition
temperature and its order.~\cite{ceperley95}

The area estimator of Eq.~(\ref{eq:area_estimator}) gives a scalar
value for the global superfluid fraction
$\rho_s/\rho$.\index{superfluidity!fraction} This provides a complete
description for homogeneous helium systems.  However, a finite cluster
of nanoscale dimensions necessarily contains inhomogeneity deriving
from the surface, and atomic and molecular dopants provide additional
sources of inhomogeneities.  In this situation
Eq.~(\ref{eq:area_estimator}) may be interpreted as providing an
estimate of the {\em global} superfluid fraction averaged over all
sources of inhomogeneity.  It is notable that the impurity molecule
does not significantly perturb this global superfluid fraction.  For
neutral Na-doped clusters, the area estimator yields a global
superfluid fraction of about 95\%,~\cite{nakayama01} consistent with
the very weak perturbation of the density noted earlier.  For more
strongly bound systems such as He$_N$SF$_6$ and He$_N$Na$^+$, it is
found that $\rho_s$ is similarly large, approaching unity for $N>100$
at $T=0.625$~K.~\cite{kwon00,nakayama00} Thus to see a molecular
effect on superfluidity, one needs to examine the local solvation
structure on microscopic length scales.  Here, the density $\rho$ is
no longer uniform, particularly in the neighborhood of an
inhomogeneity.  Thus, the superfluid fraction $\rho_s/\rho$ is
expected to be dependent on position.  Some indirect indications of
this have also been found in analyses of helium
films.~\cite{shirron91}$^-$\nocite{adams92}\cite{zimmerli92}

A simple way to qualitatively estimate the {\em local} dependence of
superfluid character\index{superfluidity!local} is to examine the
probability $\Pi_p({\bf r})$ of a particle at a position ${\bf r}$ to
participate in an imaginary-time exchange cycle of length $p$.  As
discussed previously, Bose superfluidity is associated with the
existence of exchange cycles of long $p$.  In a pure cluster, the
single source of inhomogeneity is the cluster surface.  For a pure
cluster, $\Pi_{p\geq 6}(r)$ goes to zero as the radial distance $r$
approaches the surface.  In the presence of an impurity, the examples
discussed in Secs.~\ref{subsec:atomic} and
\ref{subsec:mol_impurity} show that an embedded molecule can
significantly modify the total density distribution $\rho({\bf r})$.
Consequently one also expects changes in the local superfluid
character.  Kwon and Whaley have systematically examined $\Pi_p({\bf
r})$ for helium clusters doped with a single SF$_6$
impurity.~\cite{kwon99} They define a local superfluid density
by\index{superfluidity!path integral estimators}
\begin{equation}
\rho_s({\bf r}) = \sum_{p\geq p'}^N \Pi_p({\bf r}) \rho({\bf r}) 
\label{eq:local_estimator}
\end{equation}
where $\rho({\bf r})$ is the total density at ${\bf r}$, and $p'$ is a
cutoff value for the permutation cycle length.  This does not account
for the tensor nature of the superfluid response, providing a
three-dimensional anisotropic representation of a scalar, that may be
viewed as an average over the set of tensorial response functions.
For a molecule with high symmetry such as a spherical top, this will
not be a serious limitation.  For linear molecules it will introduce
some uncertainty. For clusters of $N>50$, most of the polymers sampled
involve either one or two atoms ($p=1,2$) or many atoms (large $p$).
Thus in this size regime a clear cutoff exists.  For these sizes, Kwon
and Whaley used a value of $p'=6$.  For small clusters ($N<50$), a
clear distinction between short and long exchange cycles cannot be
made, which implies that in the small cluster regime a two-fluid
interpretation of the density\index{density!two-fluid decomposition
of} cannot be applied.

\begin{figure}
\begin{center}
%\scalebox{0.15}{\includegraphics{normal.eps}}
\scalebox{0.5}{\includegraphics{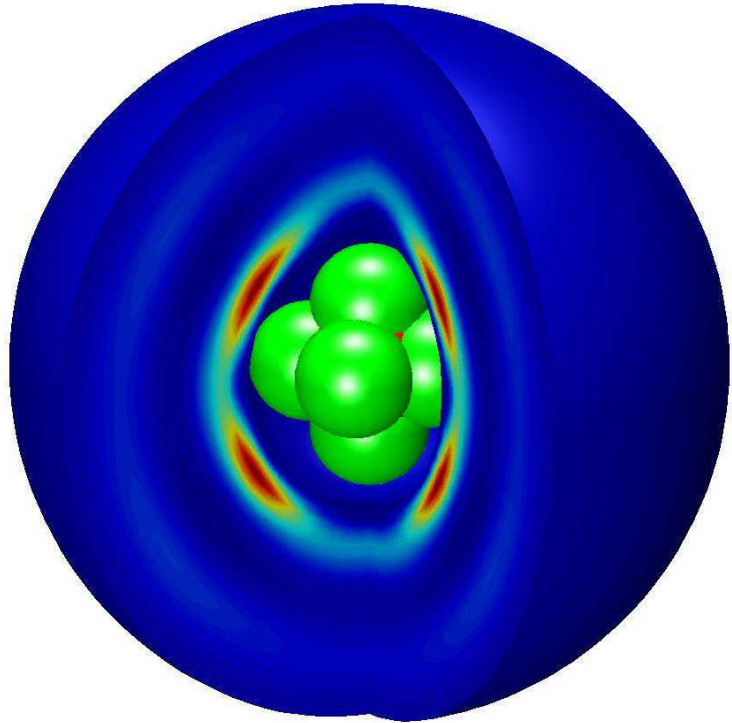}}
\end{center}
\caption[]{Local non-superfluid density $\rho_n({\bf r})$ around
SF$_6$ in a $N=64$ cluster at $T=0.312$~K, as measured by the exchange
path decomposition of the density.~\cite{kwon00} The color scale goes
from red for highest values of $\rho_n({\bf r})$, to blue for the
lowest values of $\rho_n({\bf r})$. The size of the ball corresponds
to a distance from the SF$_6$ molecule of $r=9.0$~\AA\@.  The two cuts
display the density in two equivalent planes containing $C_3$ and
$C_2$ axes.  The strong binding to the octahedral sites located along
the $C_3$ axes is evident, with 4 of the 8 octahedral sites visible
here.}
\label{fig:sf6_nsfluid}
\end{figure}

For the octahedral SF$_6$ molecule, the local superfluid estimator of
Eq.~(\ref{eq:local_estimator}) yields an anisotropic superfluid
solvation structure around the impurity molecule, whose density
modulations are similar to those of the total density $\rho({\bf r})$.
The local non-superfluid density $\rho_n({\bf r})=\rho({\bf
r})-\rho_s({\bf r})$ does depends weakly on temperature, which implies
that $\rho_n({\bf r})$ consists of thermal contribution and a
molecule-induced component.\index{molecule-induced non-superfluid}
Fig.~\ref{fig:sf6_nsfluid} shows a three-dimensional representation of
the local non-superfluid distribution around the octahedral SF$_6$
molecule.  The red areas of highest non-superfluid density are located
at the octahedral sites of strongest binding to the molecule,
reflecting the origin of this as a molecule-induced non-superfluid.
This is in contrast to the thermal normal density of bulk He~II in the
Landau two-fluid description of a homogeneous
superfluid.~\cite{landau41} The molecule-induced density component
depends on the strength and range of the helium-impurity interaction
potential, and is expected to persist at $T=0$.  Detailed analysis
shows that it is non-zero only in the first solvation layer around the
molecule.~\cite{kwon99}

\begin{figure}
\begin{center}
\scalebox{0.5}{\includegraphics{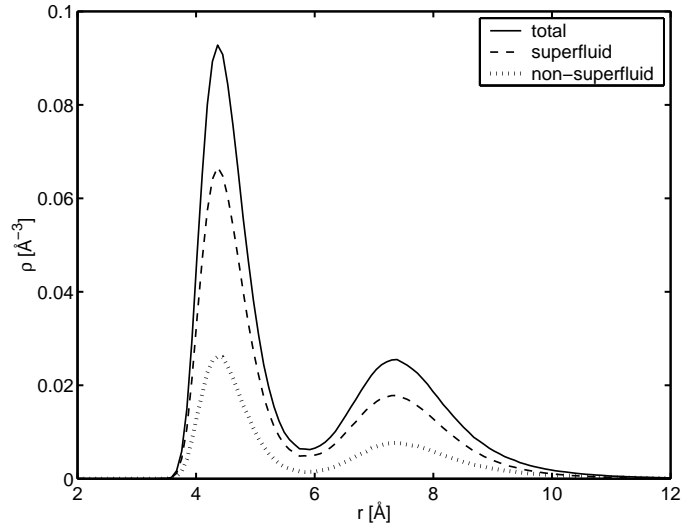}}
\end{center}
\vspace{1pc}
\caption[]{Total, local non-superfluid, and local superfluid densities
around SF$_6$ in a $N=64$ cluster at $T=0.625$~K, calculated with only
the isotropic component of the SF$_6$-He interaction potential.  The
origin is set at the impurity center-of-mass.  The local superfluid
density is calculated with the exchange path length criterion of
Ref.~\refcite{kwon99}.}
\label{fig:sf6iso}
\end{figure}

An analysis using the local estimator of
Eq.~(\ref{eq:local_estimator}) has been applied to a number of
different molecular impurities in helium clusters, including the
linear molecules OCS and HCN,~\cite{kwon00,kwon01b} and
benzene.~\cite{kwon01} These systems exhibit a similar layering in
both local superfluid density $\rho_s({\bf r})$ and local
non-superfluid density $\rho_n({\bf r})$ around the molecule.  The
non-superfluid density shows slightly stronger modulations, resulting
in a weakly anisotropic local superfluid {\em
fraction}\index{superfluidity!fraction} in addition to the component
densities themselves.~\cite{kwon00} In the more strongly bound He-OCS
case, the maximum of the non-superfluid component is roughly $\sim
50\%$ of the total density, while for the weakly bound He-HCN, the
non-superfluid, or short-exchange path, component is only $\sim 20\%$.
We note that the molecule-induced non-superfluid density is also
present around an impurity possessing an isotropic interaction with
helium, {\em i.e.}, it is not essential to have an anisotropic
interaction.  In fact the existence of a molecule-induced
non-superfluid density was first seen in calculations of the SF$_6$
molecule with isotropic interactions potentials, summarized in
Fig.~\ref{fig:sf6iso}.

Nakayama and Yamashita have pursued a similar analysis of the local
superfluid density for the He$_N$Na$^+$ cluster, which exhibits a
triple-layer structure for $N=100$.~\cite{nakayama00} While they did
not explicitly compute the local quantities $\rho_s(r)$ or $\rho_n(r)$
in their PIMC study, they observed that the helium atoms in the first
solvation shell ($r<4$~\AA) rarely participate in long exchanges.
This observation, combined with the pair distribution functions
computed with respect to atoms in the first shell, led them to
conclude that the first shell is solid-like.

As discussed previously, an even more anisotropic impurity-helium
interaction potential is provided by the benzene molecule.  For the
He$_{39}$-benzene cluster at $T=0.625$~K, the two atoms corresponding
to the two total density maxima localized at the two global potential
minima, undergo less than 2\% permutation exchanges with the
surrounding helium.  This implies that they are effectively removed
from the superfluid, {\em i.e.}, constitute a true ``dead'' adsorbed
pair of atoms.~\cite{shirron91}$^-$\nocite{adams92}\cite{zimmerli92}
This near-complete removal of individual helium atoms in the solvation
shell from participation in permutation exchanges of nearby helium
atoms has not been seen for other molecules to date. It provides an
extreme case of the local non-superfluid density, $\rho_n({\bf r})$,
in which there is no longer any partial exchange of helium atoms
between the non-superfluid and superfluid densities.  These features
of the helium solvation around a benzene molecule are expected to
appear also in clusters containing larger polyaromatic molecules such
as tetracene and naphthalene.  A systematic analysis of the effect in
planar aromatic molecules of increasing size, making the transition
from a molecular to a micron-scale substrate, would be very useful.
 
Recently another local estimator of superfluidity has been proposed
that decomposes the projected area into contributions from each local
density bin.~\cite{draeger01} This decomposition allows the anisotropy
of the response tensor to be evaluated explicitly.  Application of
this local estimator to the linear HCN trimer embedded in helium
droplets has confirmed that the superfluid density is reduced in the
first solvation layer, consistent with the presence of a local
non-superfluid density induced by the molecule-helium interaction, as
first established by Kwon and Whaley.~\cite{kwon99} Furthermore, this
new estimator shows that there is an asymmetry between the helium
response to rotation about the molecular axis, versus rotation about
an axis perpendicular to the molecular axis.  Draeger {\em et al.}\
find that the superfluid response is reduced more for rotation about
the perpendicular axes than for rotation about the molecular
axis.~\cite{draeger01} In both cases it is less than unity, implying
that there is a non-superfluid component both when rotation is
accompanied by variation in potential energy, and when there is no
variation in potential energy.  This finding supports the existence of
a local non-superfluid induced by an isotropic helium-impurity
interaction, using the exchange path analysis of Kwon and Whaley
(Fig.~\ref{fig:sf6iso}).  Thus the local non-superfluid is not
dependent on the presence of anisotropy, but derives primarily from
the stronger attraction of helium to the molecule than to itself.

These studies of various molecules embedded in He$_N$ clusters
employing different estimators of local superfluidity all point to the
existence of a molecule-induced non-superfluid
density\index{molecule-induced non-superfluid} in the first solvation
shell around a molecule.  While the details of this non-superfluid
density may be somewhat dependent on how it is defined, it is evident
from the studies of OCS, benzene, and HCN polymers made to date, that
this local non-superfluid component is a general phenomenon to be
expected for all heavy molecules.  It therefore appears to be one of
the defining features of quantum solvation in a superfluid.  The
extent of exchange between non-superfluid and superfluid densities
exhibits a dependence on the strength of the helium interaction with
the molecule.  Benzene provides an interesting extreme case of
negligible exchange between non-superfluid and superfluid density
components, while less anisotropic molecules such as SF$_6$ still
possess considerable exchange between local non-superfluid and local
superfluid.  Thus, both the interaction strength with the molecular
impurity and the symmetry of this interaction are important.  The
benzene example indicates that there are useful analogies with the
well-known ``dead'' or ``inert'' layer of helium adsorbed into bulk
solid surfaces, which will be valuable to pursue in future studies.

\section{PIMC and the connection to cluster spectroscopy}\index{impurities in He!spectroscopy of}
\label{sec:spectra}
\setcounter{equation}{0}

\subsection{Electronic spectra in He$_N$}
\label{subsec:el_spect}

Calculations of electronic spectra typically require accurate
potential energy surfaces for both ground and excited electronic
states.  This is particularly challenging for excitations in condensed
phases.  To date, theoretical work in this area has been limited to
relatively to simple systems, where the helium-impurity ground and
excited state pair potentials can be computed to good accuracy using
standard {\em ab initio} electronic structure methods.
Thermally-averaged electronic absorption spectra for the $^2P
\leftarrow\, ^2S$ transition have been computed for neutral alkali
impurities at $T=0.5$~K,~\cite{nakayama01} using a modification of the
semi-classical Frank-Condon expression for the electronic
lineshape,~\cite{lax52,cheng96}
\begin{equation}
I(\omega) \propto |M|^2 \int dR\, \rho(R,R;\beta)\, \delta[V_e(R)-E_g(R)-\hbar\omega], \label{eq:fclineshape}
\end{equation}
where $M$ is the electronic transition dipole moment, and $V_e$ is the
potential in the electronic excited state.  The quantity $E_g$ is a
local ground state energy, which is assumed to take the form
\begin{equation}
E_g(R) = T_{\mathrm{imp}}(R) + \sum_{i=1}^N V_{\mathrm{He-imp}}({\bf r}_i) + \sum_{i<j}^N V_{\mathrm{He-He}}(r_{ij}),
\end{equation}
and explicitly incorporates the kinetic energy of the impurity atom
$T_{\mathrm{imp}}$.  The terms $V_{\mathrm{He-imp}}$ and
$V_{\mathrm{He-He}}$ correspond to the helium-impurity and
helium-helium ground state pair potentials, respectively.  The
electronic excited state potential $V_e$ is obtained~\cite{nakayama01}
from the diatomics-in-molecules (DIM) model,
\begin{equation}
V_e(R) = V_{\mathrm{He-imp}}^e(R) + \sum_{i<j}^N V_{\mathrm{He-He}}(r_{ij}),
\end{equation}
where the first term $V_{\mathrm{He-imp}}^e$ is the adiabatic energies
of the alkali atom in the $^2P$ manifold interacting with the $N$
helium atoms, and the remaining helium-helium pair potentials
$V_{\mathrm{He-He}}$ are taken to be identical to the ground state.
Thus, the thermal absorption profile $I(\omega)$ can be computed by
sampling this energy difference of Eq.~(\ref{eq:fclineshape}) from a
PIMC simulation.  As discussed in Sec.~\ref{subsec:atomic}, the
neutral alkali impurities reside on the droplet surface, and the
resulting perturbation on droplet properties is weak.  The electronic
lineshape is therefore most sensitive to the details of the surface
structure near the alkali atom.  The PIMC calculations for neutral Li,
Na, and K on helium clusters of size $N=100-300$ give good qualitative
agreement with experiment.  The doublet structure ($^2P,\,^2P_{3/2}
\leftarrow\, ^2S_{1/2}$) observed in the experimental spectra for Na
and K on helium droplets~\cite{stienkemeier96} can be seen in the PIMC
calculations.  However, while both experiment and theory show that
these transitions are shifted to the blue relative to the experimental
gas phase values, the absolute value of these shifts is in general
much more difficult to obtain from theory.  Due to weak spin-orbit
coupling for Li, the doublet splitting is small relative to the
linewidth, and thus also difficult to resolve.  The PIMC absorption
spectra for the He$_N$Li system exhibits a weak red shift and a long
tail towards the blue, both of which are consistent with the
experimental spectra.~\cite{stienkemeier96}

\subsection{Vibrational shifts in infrared spectroscopy of molecules in
He$_N$}
\label{subsec:vib_spect}

The first spectroscopic experiment made on a doped helium cluster
measured the infrared absorption spectrum of the octahedral SF$_6$
molecule.~\cite{goyal92} This low resolution spectrum, obtained with a
pulsed CO$_2$ laser, revealed that the $\nu_3$ vibrations of SF$_6$
molecule are red-shifted from the gas phase value by about
$1-2$~cm$^{-1}$. They also appeared to show that the three-fold
degenerate absorption for these vibrational modes is split into two
peaks.  The split peaks were interpreted as implying that the SF$_6$
molecule resides on the cluster surface where the three-fold
degeneracy would be expected to be split into parallel and
perpendicular modes.  However, DMC calculations made at that time
showed that the molecule should be located at the cluster
center.~\cite{barnett93} This was later confirmed by PIMC
calculations~\cite{kwon96} and verified by subsequent experimental
investigations.  These include high resolution spectra made with diode
lasers~\cite{hartmann95} which showed a single vibrational absorption,
red-shifted by $\Delta\nu = -1.6$~cm$^{-1}$ and having no splitting of
the vibrational degeneracy, and analysis of ionization products of
SF$_6$-doped helium clusters.~\cite{scheidemann93} Calculations of the
spectral shifts of these triply degenerate intramolecular $\nu_3$
vibrations of SF$_6$ were made with both PIMC and
DMC,~\cite{kwon96,barnett93} using the instantaneous dipole-induced
dipole (IDID) mechanism originally proposed by Eichenauer and
Le~Roy~\cite{eichenauer88} to calculate the vibrational spectra of
SF$_6$ inside argon clusters.  PIMC allows the calculation of the
thermally averaged spectral shift at finite temperatures, while DMC
gives the ground state, $T=0$~K value of the spectral shift.
Ref.~\refcite{kwon96} provides a discussion of the difficulties in
calculating spectral line shapes (and hence extracting line widths)
from a finite-temperature path integral calculation.  The IDID
approach is taken because the intramolecular vibrational dependence of
the He-SF$_6$ interaction potential is not known, and it is therefore
necessary to approximate this.  In the IDID model, the origin of the
spectral shift is assumed to be the dipole-dipole interaction between
the instantaneous dipole moment of the SF$_6$ $\nu_3$ vibration and
the induced dipole moments of the surrounding helium atoms.  The
average shift of the $\nu_3$ absorption estimated from the IDID model
within PIMC calculations are red-shifted, in agreement with
experiment, but the magnitudes of the calculated shift ($\Delta\nu =
-0.84$~cm$^{-1}$) is somewhat smaller than the experimental value of
$\Delta\nu = -1.6$~cm$^{-1}$.  Overall, the agreement of the spectral
shift value to within a factor of two is quite reasonable, but it is
evident that for a proper understanding of the spectral shift of
SF$_6$ inside helium, one needs to also incorporate the contribution
from the repulsive part of the He-SF$_6$ interaction that is neglected
in the IDID model.

Recently Gianturco and Paesani have calculated vibrationally adiabatic
He-OCS potentials for various internal vibrational states of OCS,
which allow for a more fundamental approach to the calculation of
vibrational shifts.~\cite{gianturco00} These vibrationally adiabatic
potentials are derived by evaluating the interaction potentials as a
function of both external coordinates and an internal vibrational
coordinate using {\em ab initio} electronic structure methods, and
then averaging over the internal vibrational wavefunctions.  In
particular, they have provided vibrationally adiabatic potentials
$V_{00}$ and $V_{11}$ that are averaged over the ground state and
first excited state of the asymmetric stretching motion of the
molecule, respectively.  The shift of an intramolecular vibrational
mode inside the helium cluster can be estimated within an adiabatic
separation of the fast intramolecular vibrational mode from the slow
He-He and He-molecule degrees of freedom.~\cite{blume96} In this
approach, the spectral shift results from computing the average of the
difference between $V_{00}$ and $V_{11}$ over the finite-temperature
ensemble sampled in the PIMC simulation.  Recent PIMC calculations for
He$_{39}$-OCS at $T=0.3$~K find a red-shifted asymmetric vibration,
with the shift of $\Delta\nu = -0.87(1)$~cm$^{-1}$.~\cite{paesani01}
The sign of the shift is in agreement with that seen in experimental
measurements for OCS made in larger clusters involving more than 1000
helium atoms at $T=0.38$~K, but its magnitude is somewhat larger than
the experimental value of $\Delta\nu =
-0.557(1)$~cm$^{-1}$.~\cite{grebenev00c} Detailed analysis indicates
that these discrepancies are likely due to small errors in the
vibrationally averaged adiabatic potentials.~\cite{paesani01}

\subsection{Rotational spectra of molecules embedded in He$_N$}
\label{subsec:rot_spect}

The experimental observation of rotational fine structure for infrared
spectra of vibrational transitions in the bosonic $^4$He clusters but
not in the corresponding fermionic $^3$He clusters at the operative
temperature of $T=0.38$~K,~\cite{grebenev98} led to the conclusion
that superfluidity is essential for observation of a free rotor-like
spectrum.  This has been explained as a result of the weak coupling of
molecular rotations to the collective excitations of superfluid He~II,
compared to the much stronger coupling to particle-hole excitations in
the Fermi fluid $^3$He.~\cite{babichenko99} Consequently, the
rotational lines are considerably broadened in the fermionic clusters,
and the fine structure of rovibrational transitions is washed out.
This is consistent with the results of direct calculations of
rotational energy levels of $^4$He clusters containing rotationally
excited molecules, using zero-temperature DMC-based
methods.~\cite{kwon00,blume99,viel01b,lee99,paesani02} These direct
calculations show that the bosonic nature of the $^4$He is critical in
ensuring a free rotor-like spectrum of rotational energy levels of the
molecule when embedded in a helium droplet.  The corresponding
rotational energy levels in fermionic helium droplets have not yet
been calculated, and would constitute an interesting theoretical topic
for future study.  Path integral calculations have not provided any
information on the dynamical differences resulting from solvation in
fermionic versus bosonic helium droplets so far, since fermionic PIMC
simulations have not been made for these systems.

A major feature of the rotational spectra of molecules in $^4$He
droplets is the appearance of free rotor-like spectra with increased
effective moments of inertia. In principle, any helium-induced change
in the molecular moment of inertia should be directly related to a
change in the global superfluid fraction, according to
Eqs.~(\ref{eq:frac_I}) and (\ref{eq:area_estimator}) (assuming that
linear response measures are applicable to the quantized rotation of
the molecule).  Furthermore, a highly anisotropic molecule would be
expected to result in some anisotropy in the helium response for
rotation around different axes, yielding anisotropy in the tensor of
global superfluid response.~\cite{kwon01} However, as noted earlier,
the global superfluid estimator is relatively insensitive to the
presence of an impurity and the statistical errors mask small changes.
It is possible that for significantly larger, and more anisotropic
molecules than those theoretically studied to date, {\em e.g.}, for
the planar aromatic molecules such as tetracene and phthalocyanine
that have already been studied
experimentally,~\cite{hartmann97}$^-$\nocite{hartmann98}\cite{hartmann02}
the global superfluid response may be more affected and yield
information.  For the relatively small molecules and complexes studied
so far however, it has proven necessary to examine the local
perturbations of the helium superfluidity in order to develop an
understanding of the coupling between this and the molecular
rotational dynamics.

The microscopic two-fluid description\index{two-fluid theory} of the
quantum solvation of molecules in He that is provided by path integral
calculations has led to a detailed analysis of the effective moments
of inertia of molecules solvated in a bosonic superfluid, and hence to
a quantitative understanding of the effective rotational constants
measured in the infrared and microwave spectroscopy experiments.
Since the path integral calculations carried out to date do not
explicitly incorporate the molecular rotational degrees of freedom,
the connection between the path integral densities and the molecular
moments of inertia has to be made within a dynamical model.  Kwon and
Whaley have proposed a quantum two-fluid model for calculating the
effective moment of inertia.~\cite{kwon00,kwon99} The main features of
this model are summarized below.  As will be evident from the
assumptions of this quantum two-fluid model for superfluid helium
response to molecular rotation, it is applicable only to the regime of
heavier molecules, {\it i.e.}, those possessing gas phase rotational
constants less than $\sim 0.5$~cm$^{-1}$.  Excellent agreement with
spectroscopic measurements is obtained for the two instances in which
the He-molecule interaction potential is best known, SF$_6$ and
OCS.~\cite{kwon00} The theoretical values of rotational constant
calculated from the quantum two-fluid model are 0.033~cm$^{-1}$ and
0.067~cm$^{-1}$, for SF$_6$ and OCS respectively, compared with the
corresponding experimental values 0.034(1)~cm$^{-1}$ and
0.073~cm$^{-1}$.  Draeger {\em et al.}\ have recently tested this
quantum two-fluid model for the linear trimer
(HCN)$_3$.~\cite{draeger01} They also find excellent agreement between
the predictions of the quantum two-fluid model and the experimentally
measured rotational constant.  While the HCN monomer lies in the
regime of light molecules, the trimer is sufficiently massive to fall
within the heavy regime, possessing a gas phase rotational constant of
$B_0=0.015$~cm$^{-1}$.~\cite{miller88}

The quantum two-fluid model of Kwon and Whaley is a microscopic
two-fluid continuum theory for the spectroscopic response of a
molecule rotating in superfluid He.  It is to be distinguished from
the phenomenological two-fluid theory of Landau for bulk
He~II\@.~\cite{landau41} The Kwon and Whaley model examines the
helium response to rotation of an embedded molecule that, starting
from the local two-fluid decomposition of the molecular solvation
density into non-superfluid and superfluid components.  Unlike the
Landau two-fluid theory for bulk He~II, it does not make a two-fluid
decomposition of the current densities, but deals only with
decomposition of the helium density near an impurity on an atomic
length scale, into a non-superfluid component induced by the molecular
interaction and the remaining superfluid.  This constitutes a
significant difference between the well-known phenomenological theory
for bulk, homogeneous He~II, and the microscopic two-fluid model for
molecular rotational dynamics in an inhomogeneous superfluid solvation
situation.  For the remainder of this section we shall interchangeably
use the terms two-fluid model, microscopic two-fluid model, and
quantum two-fluid model to refer to the Kwon/Whaley model.

The starting point for the quantum two-fluid model for helium response
to molecular rotation is the local two-fluid density decomposition of
the molecular solvation density that results from path integral
calculations.  As described in Sec.~\ref{subsec:nsfluid}, consistent
evidence for the existence of the local non-superfluid density in the
first solvation shell around the molecule, induced by a strong
molecular interaction with helium, has now been obtained from two
different estimators of the local superfluid response.  The second
feature of the model is the assumption of adiabatic
following\index{adiabatic following} of some or all of the solvating
helium density with the molecular rotation.  Adiabatic following means
that the helium density follows the molecular rotational motion.
Quantitatively, complete adiabatic following of the helium density
would imply that when viewed in the rotating molecular frame, the
helium density appears stationary.  Thus, in the molecular frame it is
independent of rotational state.  This applies to both classical and
quantum descriptions of the molecular rotation.  In a classical
description the helium density is analyzed as a function of continuous
molecular rotation frequency, while for a quantum description it is
analyzed as a function of quantum rotational state of the molecule.

The accuracy of the adiabatic following assumption, as well as
quantification of the extent of adiabatic following by helium, has
been the subject of several studies by Whaley and
co-workers.~\cite{kwon00,lee99,patel01b} Within a classical
description of molecular rotation, Kwon {\em et al.}\ provided a
criterion for adiabatic following, namely that the kinetic energy of
rotation associated with a particular helium density (total,
non-superfluid, or superfluid) be less than the potential energy
barrier to rotation around the molecule.~\cite{kwon00} This criterion
is applicable to densities deriving from any number of helium atoms,
and allows simple estimates using either barriers to rigid rotation,
or barriers to adiabatic motion between potential minima associated
with different molecular orientations.  Application of this criterion
to the molecules OCS, SF$_6$ and HCN, for which the molecule-helium
interaction potentials are very well characterized, showed that for
both OCS and SF$_6$ it is energetically feasible for the entire helium
density to adiabatically follow the molecular rotation.  However, for
the lighter HCN molecule, it is not energetically feasible for any
density component, whether non-superfluid, superfluid, or total, to
adiabatically follow the molecular rotation.  In a classical sense,
the molecular rotation is then too fast for the helium to follow.  The
consequence of this lack of adiabatic following is that the helium
density distribution is more diffuse when viewed in the molecular
frame.  This cannot be seen directly in the PIMC densities, since
molecular rotation is not included in these.  However, it can be seen
directly in diffusion Monte Carlo calculations of excited rotational
states of the molecules in He clusters.\index{Monte Carlo!diffusion}
In these calculations, made with an importance sampling algorithm for
rotational degrees of freedom,~\cite{viel01} the helium density (or
wave function) is projected into the rotating molecular frame and
compared with the corresponding density (wave function) from a
calculation performed without molecular rotation.~\cite{kwon00,lee99}
Explicit analysis of the dependence on rotational state can also be
performed, although comparison between rotating and non-rotating cases
is already very revealing.  The original application of this analysis
showed that the extent of adiabatic following decreases for lighter
molecules, with the helium density in the molecular frame becoming
more diffuse as the rotational constant of the molecule
increases.~\cite{lee99} Kwon {\em et al.}\ showed recently how this
comparison may be quantified by evaluation of a quality factor $Q$
that measures changes in the ratio of densities along directions
corresponding to strong and weak binding, as a function of molecular
rotational state.~\cite{kwon00} Complete adiabatic following is
measured by $Q=1$, provided the molecular interaction potential is
anisotropic.  (For an isotropic interaction with helium, adiabatic
following is not applicable, and $Q\equiv 1$ by definition.)
Application to the series of molecules, OCS, SF$_6$, and HCN, shows
that $Q \sim 0$ for HCN, and $Q \sim 0.7$ for both OCS and
SF$_6$.~\cite{patel01b} This confirms the prediction of the energetic
criterion for HCN, {\it i.e.}, there is negligible adiabatic following
around this molecule.  The $Q$-value results for the heavier molecules
are quite significant, implying that the extent of adiabatic following
is not complete, even for the most strongly bound case of a single He
atom attached to SF$_6$.~\cite{kwon00} So only a {\em fraction} of the
helium density can adiabatically follow the molecular rotation, even
for a heavy, strongly bound molecule.

The next stage of the quantum two-fluid model is to consider the
consequences of adiabatic following for both the local non-superfluid
and local superfluid density around a dopant molecule.  These two
density components show very different response to adiabatic
following, deriving essentially from the different spatial extent that
results from their corresponding underlying exchange permutation
paths.  The molecule-induced non-superfluid
density\index{molecule-induced non-superfluid} is localized close to
the molecule, within the first solvation layer, and is composed of
very short permutation exchange paths.  In order to satisfy adiabatic
following, such a localized density must rotate rigidly with the
molecule.  There is no other obvious way in which a density that is
spatially localized within a few angstrom can remain constant in a
rotating molecular frame.  This results in an increment of moment of
inertia from the local, molecule-induced non-superfluid that is given
by
\begin{equation}
I_{n} = m_4\int_V d{\bf r}\, \rho_{n}({\bf r}) r_{\perp}^2,
\end{equation}
\label{eq:I_nsfluid}
For the heavy molecules SF$_6$, OCS, and the linear trimer (HCN)$_3$,
the PIMC values for $\Delta I_{n}$ amount to 100\%, 90\%, and $\sim
81\%$ of the corresponding experimentally observed moment of inertia
increments, $\Delta I$.  It is interesting that for the highly
symmetric SF$_6$ molecule, a very similar result ($\Delta I_{n}\sim
98\%$ of $\Delta I$) is obtained from calculations with only an
isotropic interaction potential.  While there is no adiabatic
following with an isotropic interaction and hence no mechanism for
rigid coupling of the non-superfluid helium density to the molecular
rotation, the high symmetry of the octahedral SF$_6$ molecule
nevertheless results in the integrated non-superfluid density in the
first shell being very similar in anisotropic and isotropic
calculations.  In fact, the finding that the isotropic non-superfluid
density could account quantitatively for $\Delta I$ was obtained prior
to calculations of the anisotropic local non-superfluid
density.~\cite{kwon97} While this result did not have the theoretical
justification of rigid coupling as a result of adiabatic following at
that time, it was the first indication that a local two-fluid
description was dynamically relevant and prompted the application of
an microscopic Andronikashvili analysis of experimental rotational
spectra for the case of OCS in He.~\cite{grebenev98}

In contrast to the local non-superfluid density, the superfluid
density, while also modulated around the molecule, is not restricted
on an angstrom length scale within the quantum solvation structure.
By its very definition, consisting of long exchange paths, the
superfluid density extends far from the molecule.  Thus the equation
of continuity can applied to this density over long distances.  Kwon
{\em et al.}\ have shown that for a classical molecular rotation,
determined by a continuous frequency ${\bf\omega}$, the condition of
adiabatic following, if satisfied, can be combined with the equation
of continuity to eliminate the explicit time dependence of the density
and to arrive at an equation for the superfluid
velocity:~\cite{kwon00}
\begin{equation}
{\bf\nabla}\cdot[\rho_s({\bf r},t){\bf v}_s({\bf r},t)] = {\bf \nabla}\rho_s({\bf
r},t)\cdot({\omega}\times{\bf r}). 
\end{equation}
\label{eq:hydro_v}
The irrotational nature of a superfluid may be used to replace ${\bf
v}_s$ by $(\hbar/m_4)\nabla u({\bf r},t)$, to arrive at a second-order
partial differential equation for the superfluid velocity potential
$u({\bf r})$:
\begin{equation}
{\bf\nabla}\cdot[\rho_s({\bf r}){\bf\nabla}u({\bf r})] = \left(\frac{m_4}{\hbar}\right){\bf\nabla}\rho_s({\bf r})\cdot({\omega}\times{\bf r}). \label{eq:hydro_u}
\end{equation}
This equation, discussed in detail in Ref.~\refcite{kwon00}, was first
proposed in 1997 before full anisotropic superfluid densities in three
dimensions were calculated.~\cite{sartakov97}

A similar equation was recently presented by Callegari and
co-workers,~\cite{callegari99} together with the somewhat different
assumption that the entire local solvation density is superfluid.
Solution of these hydrodynamic equations leads to a hydrodynamic
moment of inertia increment $\Delta I_h$ that is derived from the
excess fluid kinetic energy associated with the flow pattern of ${\bf
v}_s$.  Callegari {\em et al.}\ solved the hydrodynamic equations for
several linear or rod-like molecules for which the equations become
two-dimensional, using total densities derived from density functional
calculations.~\cite{callegari99} The full solution for a molecule
showing true three-dimensional anisotropy was made recently for SF$_6$
using PIMC densities.~\cite{kwon00} Draeger {\em et al.}\ have applied
the hydrodynamic treatment to the (HCN)$_3$ trimer, using their PIMC
densities and also assuming the total density to be superfluid, for
the purpose of comparison.  It appears that very different results are
obtained for different molecules within the hydrodynamic treatment.
For octahedral SF$_6$, the value of $\Delta I_h$ is small,
irrespective of whether the total density or superfluid density is
used as input to the hydrodynamic calculations (6\% and 9\% of $\Delta
I$, respectively).  For the linear (HCN)$_3$ trimer, Draeger {\em et
al.}\ find an upper bound of $\Delta I_h \sim 0.7\Delta I$, when the
total density is assumed superfluid ($\Delta I_h = 850$~amu~\AA$^2$,
compared with an experimental value~\cite{miller88} of $\Delta I =
1240$~amu~\AA$^2$).  The calculations of Callegari {\em et al.}\ for
rod-like molecules yielded between $\Delta I_h \sim 67$\% and 98\% of
the experimentally measured increments $\Delta I$.  These studies
differed from those for SF$_6$ and (HCN)$_3$ in that input densities
were obtained from density functional calculations rather than from
PIMC, in some cases using simple estimates from pairing rules to
construct interaction potentials when no empirical or {\em ab inito}
potentials were available.

The hydrodynamic treatment of the local superfluid density derived
from PIMC has a number of questionable aspects.~\cite{kwon00} Firstly,
the treatment of the molecular rotation as a classical rotation
characterized by a continuous frequency $\omega$ must be reconciled
with the intrinsic quantized nature of spectroscopic transitions
between quantum rotational states.  The response to classical rotation
necessarily gives rise to angular momentum generation in the
superfluid, analogous to the rotation of bulk superfluid in a
superleak.~\cite{mehl65,kojima71} Kwon {\em et al.}\ have calculated
the angular momentum generation by absorption of a photon within a
semiclassical analysis, and shown that significant values of $\Delta
I_h$ result in large fractions of the photon angular momentum being
transferred to the superfluid density component.  This contradicts
conclusions of a number of zero-temperature DMC-based calculations
that indicate there is negligible transfer of angular momentum to the
fluid on rotational excitation.~\cite{blume99,viel01,lee99} Kwon {\em
et al.}\ resolved this by adding quantum constraints to the
hydrodynamic formulation, and concluding that violation of these
indicates invalidity of the hydrodynamic contribution.  This in turn
may derive from lack of complete adiabatic following, for which
considerable evidence now exists, as outlined above, or from the
intrinsic lack of applicability of hydrodynamics to the motions of a
superfluid on the atomic length scale.  An indicator of this breakdown
is the fact that the solutions to the hydrodynamic equations with
density inputs of atomically modulated helium solvation densities
around an embedded molecule, show variations over length scales of 1
to 2~\AA\@.~\cite{kwon00,callegari99} Such variations on a distance
comparable to or less than the coherence length $\xi$ of helium imply
that a hydrodynamic solution is at its limits of validity here, at
best, and should be interpreted with great caution.

The overall conclusions of the quantum two-fluid model for the
response of helium to rotation of an embedded molecule are thus that
the primary contribution to the increased molecular moment of inertia
is a rigid coupling to the local non-superfluid density in the first
solvation shell.  This yields 100\%, 90\%, and $\sim 81\%$ of $\Delta
I$ for the heavy molecules SF$_6$, OCS, and (HCN)$_3$, respectively.
The accuracy of these estimates is dependent on the accuracy of the
underlying molecule-helium interaction potentials.  In contrast, there
appears to be negligible contribution from the superfluid, whose
response must be restricted by angular momentum constraints.  This is
consistent with the findings of only partial adiabatic following of
the total helium density.  It appears reasonable that only the
non-superfluid density adiabatically follows the molecular rotation,
while the superfluid density, which is defined over much longer length
scales, cannot effectively adiabatically follow.  For heavy molecules,
this two-fluid model provides a complete dynamical picture.

For light molecules such as HCN, the zero-temperature calculations
have shown that adiabatic following is questionable even for the
non-superfluid density. Consequently, in this situation the two-fluid
model cannot be used to estimate effective moments of inertia. At this
time, the zero-temperature DMC-based direct calculations of rotational
energy levels of doped clusters provide the only route to microscopic
theoretical understanding of spectroscopic measurements of rotational
transitions for such light molecules in helium droplets.~\cite{viel01}
This will hopefully change in the future, when molecular rotational
motions are explicitly incorporated into the PIMC.
 
\section{Conclusions and future directions}
\label{sec:conclude}
\setcounter{equation}{0}

The path integral approach has provided a powerful theoretical tool
for investigating the superfluid properties of finite helium droplets.
Path integral Monte Carlo calculations have shown that these systems
constitute nanoscale superfluids and offer a unique route to probing
the structure and response of a Bose superfluid on a microscopic
length scale.  They also provide examples of inhomogeneous superfluid
density, with the unique feature that this inhomogeneous, nanoscale
superfluid density can be probed by molecular and atomic dopants.

The microscopic calculations show that such impurities introduce a
local quantum solvation structure into the otherwise smoothly varying
helium density.  The numerical path integral Monte Carlo method has
allowed this quantum solvation structure in a superfluid to be
analyzed in terms of the boson permutation exchange properties, and
conversely, the effect of the molecular interaction on the superfluid
to be quantified.  PIMC calculations show that a strongly bound
impurity induces a non-superfluid density in the first solvation
shell, whose extent is determined by the strength of the molecular
interaction.  Similar conclusions are derived from analysis of the
permutation exchange path lengths into short, strongly localized
paths, and long, delocalized paths, and from decomposition of the
linear response estimator for global superfluidity.  The existence of
this local non-superfluid component in the solvation layer around
microscopic impurities therefore seems to be a general feature of
molecular solvation in superfluid He clusters.  Response of this local
two-fluid density to the rotation of a molecular impurity gives rise
to increments in the molecular moment of inertia, but does not
otherwise modify the effective free rotation of the molecule in the
superfluid.

These path integral studies of doped helium droplets open the way to
study of several intriguing questions.  One is the effect of the
molecular rotation on the quantum solvation structure in the
superfluid local environment.  As noted in this article, all PIMC
studies to date have not explicitly incorporated the molecular
rotational degrees of freedom.  We now know from zero-temperature
calculations of the quantum rotational excitations that the molecular
rotation does result in a smearing out of the angular anisotropy in
the quantum solvation structure.~\cite{kwon00,viel01b,paesani01} This
implies less than perfect adiabatic following of the helium density,
even in the rotational ground state.  Given the significance of the
adiabatic following assumption for models of the helium response and
hence for the analysis of rotational spectra of doped molecules,
developing a direct route to the solvation density around a rotating
molecule is highly desirable.  This can be done by incorporating the
molecular rotational kinetic energy in the path integral
representation.  A key question with spectroscopic implications is
then how the local two-fluid density decomposition is modified.  We
noted earlier that the moment of inertia increment of the
non-superfluid density around SF$_6$ is approximately independent of
the anisotropy of the interaction potential.  This suggests that even
if the two-fluid densities are modified with rotation, becoming less
anisotropic, the effective moment of inertia of the molecule in $^4$He
will be unchanged.  This remains to be verified.

A second direction departing from the analysis of molecular solvation
structure in a superfluid is the investigation of localization of
helium atoms and their removal from the superfluid state, as a
function of the binding to organic molecules of increasing size.  In
the study of benzene, the key feature responsible for the localization
phenomenon was identified as the strong and highly anisotropic
interaction of helium with the $\pi$-electron system.  Systematically
varying the extent of the $\pi$-electron system by going to larger
planar, polyaromatic molecules will allow the transition from a
nanosubstrate to a microsubstrate that begins to mimic a bulk solid
surface to be investigated.  We expect that the "inert" layers
familiar from studies of thin films of helium on graphite\index{helium
films!on graphite} will evolve from these localized atoms, but the
manner in which this happens will depend on the role of lateral
confinement and permutation exchanges in the presence of an extended
$\pi$-electron system.

A third, novel direction is provided by extension of these ideas to
nanoscale clusters of molecular hydrogen, H$_2$.  In its rotational
ground state, the H$_2$ molecule is a boson, but bulk superfluidity is
preempted by the occurrence of the triple point at $T=13.6$~K.
However, finite-size and reduced dimensionality systems are offer ways
of bypassing this solidification of hydrogen by allowing lower
densities and thereby moderating the effects of strong
interactions. Path integral calculations have already been used in
several instances in the search for a superfluid state of molecular
hydrogen.  Thus, very small finite clusters of (H$_2$)$_N$ ($N \leq
18$) have been shown with PIMC to be not only liquid-like but also to
show a limited extent of superfluidity.~\cite{sindzingre91}
Two-dimensional films of hydrogen have been shown to allow a stable
superfluid phase at low temperatures provided that an array of alkali
atoms is co-adsorbed, providing stabilization of a low density liquid
phase.~\cite{gordillo97} Given these low-dimensional antecedents, it
appears possible that a relatively small solvating layer of hydrogen
wrapped around a molecule might also show some superfluid behavior.
Path integral calculations are now in progress to examine the extent
of permutation exchanges in cycles around different axes of a linear
molecule wrapped with variable numbers of H$_2$
molecules.~\cite{kwon02} Such studies will provide microscopic
theoretical insight into the quantum dynamics underlying recent
spectroscopic experiments showing anomalies in the molecular moment of
inertia that are consistent with a partial superfluid response of the
solvating hydrogen layer.~\cite{grebenev00}

In summary, the path integral Monte Carlo approach provides a unique
tool for analysis of these degenerate quantum systems in finite
geometries and with chemically complex impurity dopants.  The insights
into nanoscale superfluid properties that have resulted, and the
interplay between physical and chemical effects afforded by
calculations on doped helium droplets offer promise of new
opportunities for analysis and manipulation of superfluid at the
microscopic level.

\section*{Acknowledgments}
We acknowledge financial support from the National Science Foundation
(KBW, CHE-9616615, CHE-0107541) and the Korea Research Foundation (YK,
2000-015-DP0125).  We thank NPACI for a generous allocation of
supercomputer time at the San Diego Supercomputer Center, and KORDIC
for a generous allocation of its supercomputer time.  PH acknowledges
the support of an Abramson Fellowship.  KBW thanks the Alexander von
Humboldt Foundation for a Senior Scientist Award, and
Prof. J.~P.~Toennies for hospitality at the Max-Planck Institut f\"ur
Str\"omungsforschung, G\"ottingen during a sabbatical year 1996-97.
We thank D.~T.~Moore for permission to produce Fig.~\ref{fig:displ},
and A.~Nakayama and K.~Yamashita for providing some of the data given
in Fig.~\ref{fig:denprofile}a.

\end{document}